\documentclass[11pt]{article}

\usepackage[margin=1in]{geometry}

\usepackage{amsthm}
\usepackage{amsfonts}
\usepackage{amssymb}
\usepackage{mathtools}
\usepackage{bm}

\usepackage{authblk}

\usepackage[authoryear]{natbib}

\usepackage{graphicx}
\usepackage[percent]{overpic}

\usepackage{booktabs}
\usepackage{array}
\usepackage{longtable}
\usepackage{tabularx}
\usepackage{makecell}
\usepackage{siunitx}

\usepackage{algorithm}
\usepackage[noend]{algpseudocode}

\usepackage{xcolor}
\usepackage{soul}

\usepackage[
    colorlinks=true,
    linkcolor=blue,
    citecolor=blue,
    urlcolor=blue
]{hyperref}

\newcolumntype{Y}{>{\raggedright\arraybackslash}X}

\newcommand{\ba}{\bm{a}}
\newcommand{\bq}{\bm{q}}

\newcommand{\bu}{\mathbf{u}}

\newcommand{\bx}{\bm{x}}

\newcommand{\bU}{\mathbf{U}}
\newcommand{\bV}{\mathbf{V}}

\newcommand{\bbeta}{\boldsymbol{\beta}}

\algrenewcommand{\algorithmicrequire}{\textbf{Input:}}
\algrenewcommand{\algorithmicensure}{\textbf{Output:}}
\algnewcommand{\StepHead}[1]{\item[\textbf{#1:}]}

\date{}
\title{
Sparse Reduced-rank Regression Methods for Spatially Misaligned Data
with Application to Spatial Transcriptomics
}

\author[1]{Zitian Wu}
\author[1]{Susmita Datta\textsuperscript{*}}
\author[1]{Arkaprava Roy\textsuperscript{*}}

\affil[1]{Department of Biostatistics, University of Florida}

\begin{document}

\maketitle
\begingroup
\renewcommand{\thefootnote}{\fnsymbol{footnote}}
\footnotetext[1]{These two authors jointly supervised this work.}
\endgroup

\begin{abstract}
Understanding the spatiotemporal dynamics of disease progression in relation
to transcriptomic profiles provides key insights into complex conditions such
as Alzheimer's disease. To enable such investigations, STARmap PLUS technology
offers joint profiling of high-resolution spatial transcriptomics and protein
detection within the same tissue section. Detailed visual and clustering-based
analyses of STARmap PLUS data by \citet{zeng_integrative_2023} provided
important insights into molecular and cellular changes associated with
Alzheimer's disease pathology. Motivated by this work, we develop a
kernel-weighted sparse reduced-rank regression framework that estimates
associations between plaque size and neighboring cell-level transcriptomic
profiles while enabling gene selection and borrowing strength across genes,
cell types, and disease stages. The proposed approach is implemented in a fully
automated manner with data-driven specification of key model components.
Through simulation studies, we demonstrate the robustness of the proposed
method and its superiority across a range of simulation scenarios. Applied to
Alzheimer's disease data, the proposed framework uncovers biologically
meaningful associations, highlighting its potential for advancing the
understanding of disease mechanisms.
\end{abstract}


\bigskip

\section{\label{sec:1} Introduction}

Alzheimer’s disease (AD) involves interacting pathological processes that vary across spatial locations, cell types, and disease stages \citep{de2016cellular,jack2018nia,jack2024revised}. Amyloid-$\beta$ plaques and hyperphosphorylated tau accumulate within a dynamic microenvironment shaped by microglia, astrocytes, oligodendrocyte-lineage cells, and neurons \citep{long2019alzheimer,hampel2021amyloid}. Spatially resolved studies in mouse and human reveal reproducible plaque-adjacent niches and layer-specific molecular patterns, linking microglial/astrocytic remodeling and tau-aligned neuronal changes to local pathology \citep{chen_spatial_2020,chen2022spatially,zeng_integrative_2023,mallach2024microglia,miyoshi2024spatial}. Motivated by these spatially organized niches, we adopt a plaque-centric formulation: we treat plaque centers as outcome locations and quantify how nearby cellular gene expression relates to the \textit{plaque size} (mean plaque radius) with effects allowed to vary across cell types.

Traditional transcriptomic designs struggle to capture this heterogeneity. Bulk RNA-seq averages across mixed cell types and loses cellular heterogeneity and the spatial context of the cells, while single-cell RNA-seq restores cellular resolution but still discards in-tissue organization. 
Spatially resolved transcriptomics (SRT) maps gene expression to spatial coordinates via two broad classes:
(i) barcoded capture arrays (Spatial Transcriptomics/Visium) that provide transcriptome‐wide profiles at multi-cell spot resolution \citep{staahl2016visualization,vickovic2019high,maynard2021transcriptome};
and (ii) imaging-based assays (e.g., MERFISH\citep{chen2015spatially}, seqFISH\citep{shah2016situ}, STARmap\citep{wang2018three}) that achieve single-cell or subcellular resolution with targeted gene panels.

Among imaging-based SRT, STARmap PLUS \citep{zeng_integrative_2023} co-registers targeted RNAs with amyloid-$\beta$ and p-tau immunostains within the same tissue sample, yielding single-cell expression aligned to plaque segmentations and centers, and enabling direct linkage between molecular readouts and neuropathology in situ.
Each tissue section contains many plaques and thousands of cells with known coordinates; plaque size varies markedly within a section, and this variability is embedded in distinct cellular neighborhoods defined by which cells lie nearby and what genes they express. This structure naturally motivates a plaque-centric analysis of local transcriptomic variation around pathology.

{\underline{The primary statistical challenges are:}} (1) spatial misalignment between plaque outcome locations and cell-level gene-expression measurement locations, (2) the high dimensionality of SRT data, (3) effect heterogeneity across collection times and cell types, i.e., regression coefficients linking plaque size to neighborhood features vary by time point and cell type, with gene-level coefficients exhibiting shared latent structure, (4) the need for a spatially structured characterization of different neighboring cells of different types on plaque size, and (5) section-to-section variation in local spatial structure and plaque-adjacent cellular composition, including differences in cell density, plaque density, tissue coverage, and cell-type enrichment around plaques.

Several analytic tools exist for analyzing SRT data. Many of these methods target important but distinct goals, including spatially variable gene detection, spatial domain identification, representation learning, multimodal integration, and spatial marker-relationship modeling. For example, SpatialDE \citep{Svensson2018SpatialDE} identifies spatially variable genes; graph-based methods such as SpaGCN \citep{Hu2021SpaGCN}, STAGATE \citep{Dong2022STAGATE}, and GraphST \citep{Long2023GraphST} learn spatial domains or low-dimensional representations by combining gene expression with spatial adjacency and, in some cases, histology; MISTy \citep{Tanevski2022MISTy} models marker relationships across multiple spatial views; and STACI \citep{Zhang2022STACI} integrates spatial transcriptomics with imaging-derived nuclear morphology through a graph-based autoencoder. These methods are useful for exploratory spatial discovery, representation learning, multimodal integration, and characterization of tissue organization. However, these methods are not designed to answer the research question posed above. Additional complexities arise because plaque size is measured at plaque locations, whereas gene expression is measured from many nearby cells whose locations are spatially misaligned with the plaque centers. Thus, our goal is not only to identify genes associated with plaque size, but also to characterize how nearby cell-level gene expression influences the spatially varying plaque-size outcome, while allowing these effects to vary across cell types and collection times.

Our target analysis differs from standard spatial regression problems because the relevant predictors for the spatially varying outcome here are measured at nearby, rather than identical, spatial locations. 
Spatial measurement-error models, including two-stage correction approaches and spatial simulation extrapolation methods, address problems in similar settings where spatially varying outcomes and covariates are observed with location mismatch or even at different spatial resolutions \citep{szpiro2013measurement,alexeeff2016spatial}. These methods are useful for correcting bias induced by spatial exposure error, but they are not designed for high-dimensional sparse regression problems, as ours is. 
Local/kernel regression and geographically weighted regression model spatial heterogeneity by weighting nearby observations more heavily, and multiscale GWR further allows different predictors to vary over different spatial scales \citep{nadaraya1964estimating,fotheringham2017multiscale,fan2018local,oshan2019mgwr}. These approaches are closer in spirit to our use of spatial neighborhoods, but they typically require the response and predictors to be represented on comparable observational units. In our setting, this would require constructing one predictor vector per plaque, for example, by aggregating nearby cell-level expression, which would collapse the original many-cells-to-one-plaque structure.
On the other hand, change-of-support and downscaler models address related aggregation or registration uncertainty across spatial supports \citep{gotway2002combining,berrocal2010spatio}, but classical formulations are not designed to estimate high-dimensional signed gene effects from individual neighboring cells while allowing those effects to vary across cell types and collection times.
Thus, existing spatial regression tools do not directly address the plaque-centric SRT problem considered here, where spatial misalignment, high-dimensional gene expression, cell-type/time-specific heterogeneity, and structured gene selection must be handled jointly.

Similar structural misalignment also appears in longitudinal studies, where responses and predictors may be observed at different time points. This setting is commonly referred to as asynchronous longitudinal regression. Existing methods \citep{cao_regression_2015,li_regression_2022,li_asynchronous_2023} use kernel-weighted estimating procedures to accommodate mismatched observation times between the response and covariates within a subject.
We follow a similar approach by introducing a kernel function to borrow information from all possible pairs of outcomes and the predictors in the \textit{misaligned spatial} setting. This particularly helps to characterize the collective effect of different cells from varying cell types on the plaque size.
An isotropic proximity kernel with \textit{tissue sample-specific} bandwidths aggregates local signals, accommodating section-to-section enrichment and sampling differences without overfitting directionality. 
Gene effects are parameterized by a \textit{low-rank} coefficient tensor over gene $\times$ cell type $\times$ time, together with \textit{gene-mode regularization} to capture shared structure and heterogeneity across samples and time points under high dimensionality. 
To achieve interpretable gene-level discovery, we apply a LASSO penalty only on the gene mode of this factorization: genes with weak signal are shrunk toward zero, while the associated cell-type and time patterns for the retained genes remain flexible.
All tuning parameters (e.g. kernel bandwidths, rank, and penalty levels) are selected in a \textit{data-driven} manner; details are provided in Section~\ref{sec:3:1}.
The resulting framework yields interpretable latent components through gene, cell-type, and time-point loading patterns, while the estimated coefficients provide signed local associations with plaque size: positive values indicate that higher expression is associated with larger plaque size, whereas negative values indicate association with smaller plaque size.
Applying to STARmap PLUS AD sections, the model identifies multigene expression patterns at 13 months involving microglial, oligodendrocyte-lineage, and astrocytic programs, that are discussed in Section~\ref{sec:5}.

Beyond Alzheimer’s disease, co-registering histopathology with SRT enables tumor-region delineation, quantification of tissue architecture and burden, and pathology-informed disease modeling \citep{ni2022spotclean,arora2023spatial,bassiouni2023spatial,xun2023reconstruction,jin2024advances}.
Related statistical questions also arise in environmental applications \citep{alexeeff2016spatial}. 
This breadth underscores the generality and translational relevance of our framework for modeling disease-related spatial outcomes from spatial molecular measurements.

The remainder of the paper is organized as follows. 
Section~\ref{sec:2} describes the STARmap PLUS data and associated exploratory analyses, which further motivate our modeling of plaque size through distance-weighted contributions from nearby cell-level gene expression. 
Section~\ref{sec:method} develops the proposed kernel-weighted regression model with sample-specific bandwidths and low-rank characterization with sparse gene-mode specification. 
Section~\ref{sec:4} evaluates the proposed method through simulations designed to mirror the real-data setting, and finally, Section~\ref {sec:5} applies the method to AD spatial transcriptomics data. Section~\ref{sec:conc} concludes with a discussion of limitations and possible extensions.

\section{\label{sec:2} Data and Exploratory Analysis}
\subsection{\label{subsec:2:1} Data}
We analyze single-cell spatial transcriptomics data integrated with amyloid-$\beta$ (A$\beta$) plaque histopathology from the TauPS2APP mouse model of Alzheimer's disease (AD) \citep{zeng_integrative_2023}. The data were generated using STARmap PLUS, an in situ single-cell spatial transcriptomics technology that measures gene expression while simultaneously imaging protein-level histopathological markers within the same tissue section. In the original experiment, STARmap PLUS profiled a targeted panel of 2,766 genes and imaged A$\beta$ plaques using X-34 staining and hyperphosphorylated tau using AT8 immunostaining. Although both A$\beta$ plaque and tau pathology are available in the dataset, the present work focuses on A$\beta$ plaque-centered spatial variation.

We use only the TauPS2APP AD-model samples, consisting of TauPS2APP brain tissue sections collected at two mouse ages, 8 months and 13 months. These ages represent two cross-sectional disease stages rather than repeated measurements from the same animal. In the TauPS2APP model, amyloid pathology begins before these ages and is actively expanding by 8 months; by 13 months, pathology is more advanced, with more severe amyloid and tau burden and elevated neuroinflammatory activity \citep{zeng_integrative_2023}. As tissue collection is terminal, each mouse contributes a tissue section at only one age. Thus, throughout the paper, we treat 8 months and 13 months as independent disease-stage samples rather than longitudinal follow-up times within subjects.

For each tissue section, every observed cell is characterized by its spatial location, annotated cell type, and gene-expression profile. Plaque locations are recorded in the same tissue coordinate system.
Each A$\beta$ plaque is represented by its spatial center and a plaque-size summary. In our analysis, plaque size is measured by the reported mean plaque radius from \citet{zeng_integrative_2023}.
Therefore, cells and plaques are spatially co-registered within each tissue section, enabling direct calculation of distances between each plaque and surrounding single cells.

\begin{figure}[h]
    \centering
    \begin{overpic}[width=\linewidth]{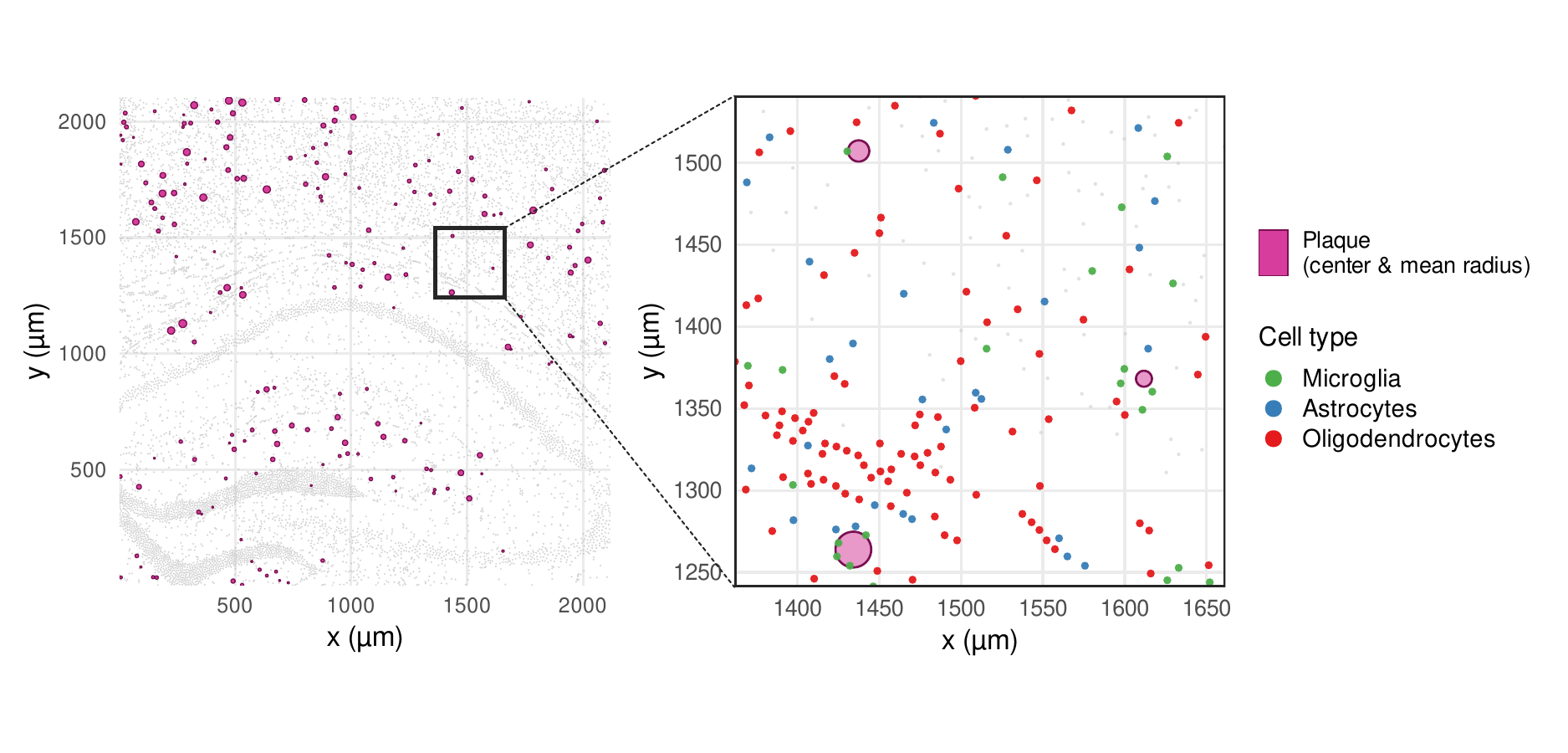}
      \put(20,2){\textbf{(a)}}\put(61,2){\textbf{(b)}}
    \end{overpic}
    \vspace{-1.8em}
    \caption{\textbf{Co-registration of single cells and amyloid-$\beta$ plaques in a representative tissue section (13 months).} 
    \textbf{(a)}~Tissue-wide map: gray points are all profiled cells, and amyloid plaques are overlaid as magenta disks centered at each
    plaque center and drawn to scale, with radius equal to the reported mean plaque radius
    (\si{\micro\meter}). The disks summarize each plaque's center and mean radius only and do
    not represent segmented plaque boundaries. The rectangle marks the region enlarged in
    (b). 
    \textbf{(b)}~Enlarged plaque
    microenvironment with three well-separated plaques: Microglia (green), Astrocytes (blue),
    and oligodendrocyte-lineage cells (red) are colored, while all other cell types are gray;
    plaques are drawn to scale as in (a). Axes are in \si{\micro\meter}. Gene expression is
    measured for each cell at its observed spatial coordinate.}
    \label{fig:coregistration}
\end{figure}

In the data, each plaque is surrounded by many single cells of different cell types, each with its own coordinates and gene-expression profile. Figure~\ref{fig:coregistration} illustrates this structure in a representative 13-month tissue section and motivates the regression framework developed later.

\subsection{\label{subsec:2:2} Exploratory spatial variation}
We first examine how cells are spatially organized around A$\beta$ plaques. For each cell, we compute the Euclidean distance from the cell location to the nearest plaque center and group cells into four distance bands: \(0\)--\(30\), \(30\)--\(70\), \(70\)--\(150\), and \(\ge150~\si{\micro\meter}\).
Since the plaque data include plaque centers but not exact plaque-boundary contours, our analyses use the plaque centers to represent plaque locations.

\begin{figure}[!htbp]
    \centering
    \includegraphics[width=0.98\linewidth]{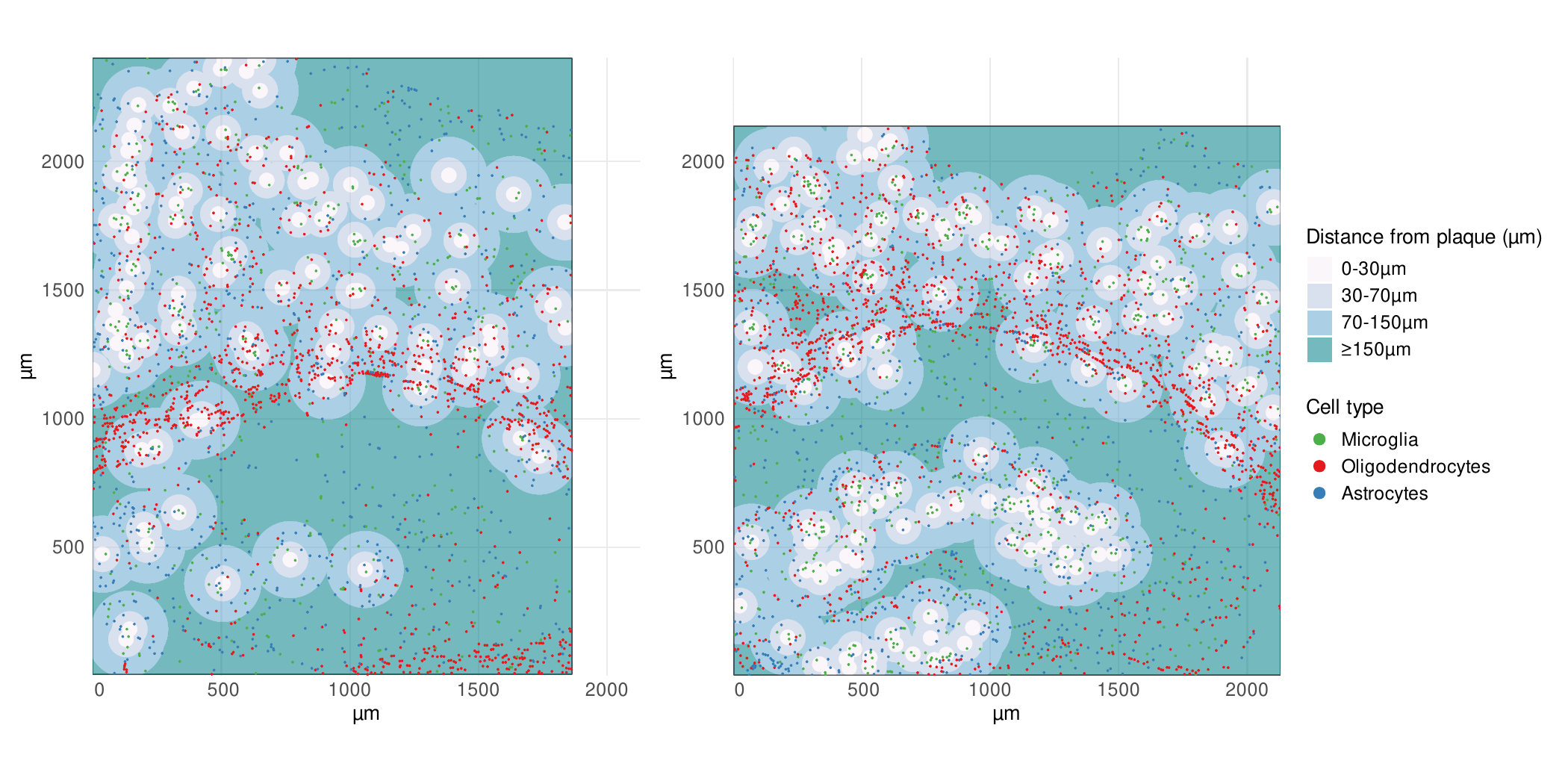}
    \caption{Plaque-centric distance maps. Shaded annuli show Euclidean distance bands from the nearest plaque center: 0--30, 30--70, 70--150, and $\ge$ 150 \textmu m. Points are cell locations colored by three cell types (legend). Each cell is placed in the ring at which it falls; these maps define the radial stratification used in downstream summaries. The two tissue sections are slightly different in dimension.}
    \label{fig:distance-plaques}
\end{figure}

\begin{figure}[!htbp]
    \centering
    \includegraphics[width=0.98\linewidth]{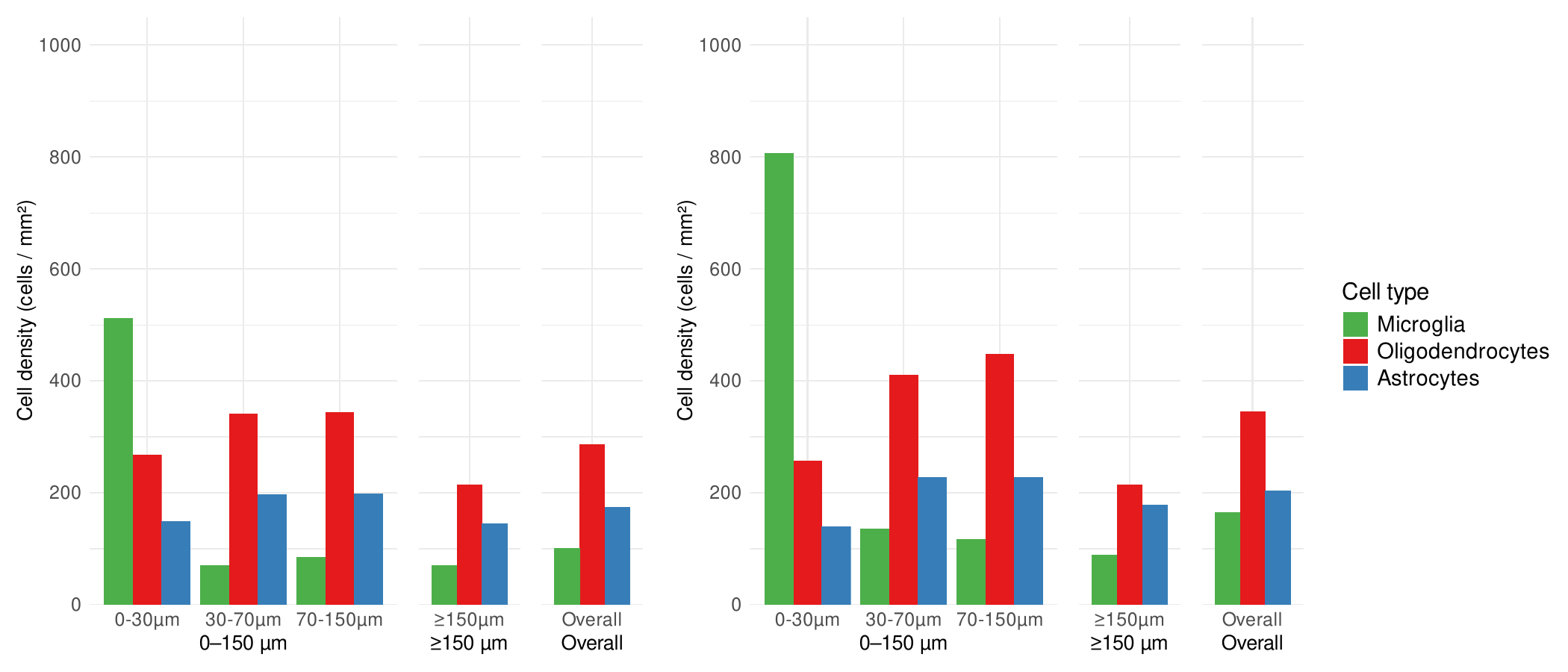}
    \caption{Cell density by distance to the nearest plaque. Bars show mean cell density (cells/mm$^{2}$) within each radial band (0--30, 30--70, 70--150, $\ge$\SI{150}{\micro\meter}) and overall, for Microglia, Astrocyte, and Oligodendrocyte-lineage cells. Densities are computed per sample and averaged across samples. Enrichment near plaques is most pronounced for Microglia in \qtyrange[range-phrase = {--}, range-units = single]{0}{30}{\micro\meter} and for Astrocyte/Oligodendrocyte-lineage cells in \qtyrange[range-phrase = {--}, range-units = single]{30}{150}{\micro\meter}.}
    \label{fig:cell-density}
\end{figure}

Guided by the plaque-associated cellular organization reported in \citet{zeng_integrative_2023} and by our exploratory summaries, we focus on three cell-type groups: Microglia, Astrocytes, and Oligodendrocyte-lineage cells. 
These cell types show clear plaque-centric spatial organization and are sufficiently represented for stratified summaries. 
Figures~\ref{fig:distance-plaques} and \ref{fig:cell-density} show that Microglia are most concentrated near plaque centers, especially within \(0\)--\(30~\si{\micro\meter}\). 
Astrocytes and Oligodendrocyte-lineage cells are less concentrated in the innermost band but show higher densities in intermediate region, approximately \(30\)--\(150~\si{\micro\meter}\) from plaque centers. These patterns suggest that plaque-associated cellular organization is strongest within approximately \(150~\si{\micro\meter}\) of plaque centers.

These plaque-centric cell-density patterns motivate a second exploratory question: whether gene expression measured in near-plaque cells varies with plaque size and disease stage. For this analysis, each cell is assigned to its nearest plaque and is considered as near-plaque if its distance to that plaque center is no greater than $\SI{150}{\micro\meter}$. Then, plaques are divided into small and large groups using the global median plaque size of $6.56$ \si{\micro\meter}, for visualization only. For each plaque and gene, we summarize local expression by averaging expression over near-plaque cells assigned to that plaque, and then compare these plaque-level summaries across plaque-size-based groups and disease stages.

\begin{figure}[!htbp]
    \centering
    \begin{overpic}[width=\linewidth]{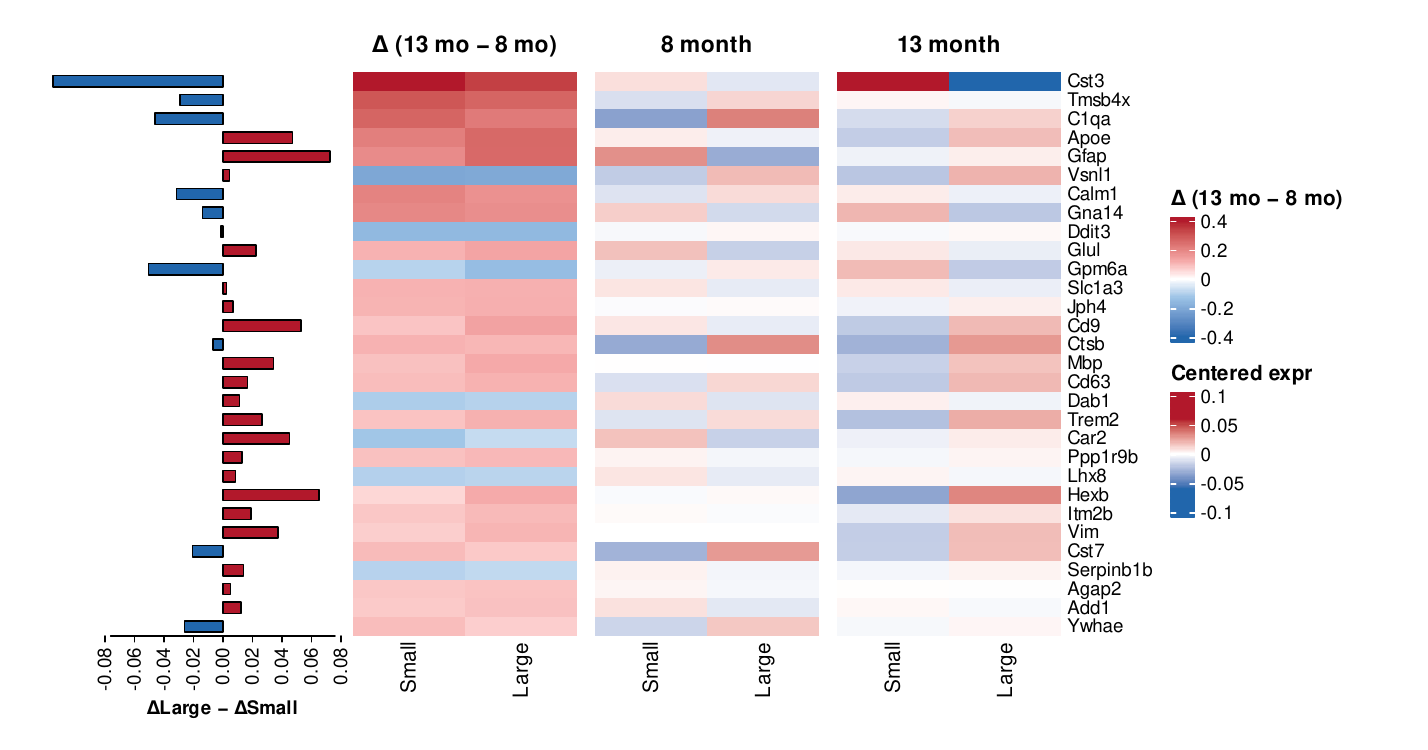}
      \put(14,-1){\textbf{(a)}}\put(31,-1){\textbf{(b)}}\put(48,-1){\textbf{(c)}}\put(65,-1){\textbf{(d)}}
    \end{overpic}
    \caption{\textbf{Near-plaque gene expression by plaque size and disease stage.}
    \textbf{(a)} Difference between the 13-month minus 8-month disease-stage contrast for large plaques and the corresponding contrast for small plaques.
    \textbf{(b)} Disease-stage contrast shown separately for small and large plaques, computed as the 13-month near-plaque expression summary minus the 8-month summary.
    \textbf{(c)} Near-plaque expression at 8 months.
    \textbf{(d)} Near-plaque expression at 13 months.
    Thus, panel (b) is computed from the underlying expression summaries in panels (c) and (d), and panel (a) is computed as the difference between the large-plaque and small-plaque contrasts in panel (b). Plaques are divided into small and large groups using the median of the reported mean plaque radii. Rows are the 30 genes with the largest disease-time-point variation. Row centering is applied only to the expression heatmaps in (c) and (d); contrast panels in (a) and (b) show raw contrasts.}
    \label{fig:time-trend-by-size}
\end{figure}

Figure~\ref{fig:time-trend-by-size} suggests that near-plaque expression patterns differ by plaque size and disease stage. These exploratory patterns motivate the regression framework below, where plaque size is modeled as a function of local cell-level gene expression.

\section{Model}
\label{sec:method}
We propose a novel kernel-weighted regression framework to analyze the effect of spatial transcriptomic profiles on amyloid-$\beta$ plaque size. For clarity, we first introduce some notations that are used throughout the article. 

We use three indices throughout the model: \(i\) indexes samples (e.g., tissue sections), \(j\) indexes outcome locations within a tissue section (e.g., plaque locations), and \(k\)
indexes predictor locations within a tissue section (e.g., cell locations). 
Let \(n\) denote the number of independent tissue sections. As each tissue section comes from a different mouse and tissue collection is terminal, we treat the tissue sections as independent cross-sectional samples.

For tissue section \(i=1,\ldots,n\), let \(M_i\) denote the number of plaques and let \(\bU_{i,j}=(U_{i,j}^{(1)},U_{i,j}^{(2)})^\top\in\mathbb{R}^2\), \(j=1,\ldots,M_i\), denote the two-dimensional spatial coordinate of plaque \(j\). 
At each plaque location, we observe a scalar outcome \(y_{i,j} = y(\bU_{i,j})\), representing plaque size.
Let \(N_i\) denote the number of cells in tissue section \(i\), and let
\(\bV_{i,k}=(V_{i,k}^{(1)},V_{i,k}^{(2)})^\top\in\mathbb{R}^2\), \(k=1,\ldots,N_i\), denote the two-dimensional spatial coordinate of cell \(k\). The spatial transcriptomic measurement at cell \(k\) is the \(p\)-dimensional gene-expression vector \(\bx_{i,k} = \bx(\bV_{i,k})\in\mathbb{R}^p\). 
Each cell has a cell-type label \(c_{i,k}\in\{1,\ldots,C\}\), and each tissue section has a collection time point \(t_i\in\{1,\ldots,T\}\).

Our goal is to estimate the association between gene-expression profiles and plaque size at different collection time points. 
For a fixed cell type \(c\) and time point \(t\), let
\(\bx^\star_{c,t}(\bu)\in\mathbb R^p\) denote the ideal local
gene-expression profile for cells of type \(c\) at spatial location \(\bu\).
This quantity is not directly observed in the spatially misaligned data, but it
represents the cell-type-specific local transcriptomic profile that would be
available in an ideal co-located study.
In such an ideal setting, the working regression model for plaque \(j\) in
tissue section \(i\) would be
\begin{equation}
    y(\bU_{i,j})
    =
    \{\bx^\star_{c,t_i}(\bU_{i,j})\}^\top \bbeta_{c,t_i}
    +
    \epsilon_{i,j}.
    \label{eq:ideal-model}
\end{equation}
Here, \(\bx^\star_{c,t_i}(\bU_{i,j})\in\mathbb{R}^p\) denotes the ideal, unobserved gene-expression profile for cells of type \(c\) at the plaque location \(\bU_{i,j}\), and \(\bbeta_{c,t_i}\in\mathbb{R}^p\) is the corresponding cell-type- and time-specific regression effect.

However, model~\eqref{eq:ideal-model} cannot be fitted directly because the gene-expression predictors are not observed at the outcome locations. Plaque size \(y_{i,j}=y(\bU_{i,j})\) is observed at plaque center \(\bU_{i,j}\), whereas cell-level gene expression is observed at cell locations \(\bV_{i,k}\), with measurements \(\bx_{i,k}=\bx(\bV_{i,k})\). In general, the plaque centers and cell locations do not coincide. Thus, the ideal predictor \(\bx^\star_{c,t_i}(\bU_{i,j})\) in model~\eqref{eq:ideal-model} is unobserved, and there is no direct one-to-one paired structure among the observation of plaque size and transcriptomic predictors at the same spatial location, such as \(\{y(\bU), \bx(\bU)\}\).

This spatial misalignment is one of several features that the proposed model must address:
\begin{enumerate}
    \item[(a)] Plaque outcomes and cell-level transcriptomic predictors are observed at different spatial locations.
    \item[(b)] Estimating cell-type-specific regression effects is of interest.
    \item[(c)] Gene-specific regression effects may share a common structure across cell types.
    \item[(d)] Samples collected at different time points are from independent biological replicates, but regression effects across time points may share a common latent biological structure.
\end{enumerate}

We treat~\eqref{eq:ideal-model} as defining, for each cell type $c$ and time point $t$, a
working association between plaque size and the local type-$c$ expression profile. The
ideal predictor $\boldsymbol{x}^{\star}_{c,t}(\mathbf{U}_{i,j})$ is unobserved; however,
if $\boldsymbol{x}^{\star}_{c,t}(\cdot)$ varies smoothly in space, the expression of a
type-$c$ cell near plaque $j$ approximates
$\boldsymbol{x}^{\star}_{c,t}(\mathbf{U}_{i,j})$, with accuracy decreasing in distance.
This motivates replacing the unavailable co-located predictor with nearby type-$c$ cells,
weighted by spatial proximity. The resulting criterion~\eqref{eq:objective} is additively
separable across cell types: each $\boldsymbol{\beta}_{c,t}$ is identified from type-$c$
plaque--cell pairs alone, so $\boldsymbol{\beta}_{c,t}$ is a kernel-weighted analogue of
the \emph{marginal} type-$c$ regression coefficient in~\eqref{eq:ideal-model}, rather than a
partial effect adjusting for other cell types. Borrowing of strength across cell types is
introduced solely through the low-rank structure, discussed later.

To address this spatial misalignment, we replace the unavailable paired regression in~\eqref{eq:ideal-model} with a kernel-weighted estimating criterion over observed plaque--cell pairs. For tissue section \(i\), plaque \(j\), and cell \(k\), we compare the observed plaque size \(y_{i,j}\) with the fitted value \(\bx_{i,k}^\top\bbeta_{c_{i,k},t_i}\) obtained from the gene-expression profile of cell \(k\). This comparison leads to considering the squared residual
\(
\left\{
y_{i,j} - \bx_{i,k}^\top \bbeta_{c_{i,k},t_i}
\right\}^2,
\)
which is weighted by
\(
K_{i,j,k}=K_{h_i}\left(\|\bU_{i,j}-\bV_{i,k}\|_2\right).
\)
The weight depends on the Euclidean distance between plaque center \(\bU_{i,j}\) and cell location \(\bV_{i,k}\), with \(h_i>0\) being a bandwidth parameter for tissue section \(i\).

The resulting kernel-weighted loss is
\begin{align}
    \mathcal L\left(\{\bbeta_{c,t}\}\right)
    =
    \frac{1}{2}
    \sum_{i=1}^n \sum_{j=1}^{M_i} \sum_{k=1}^{N_i}
    K_{i,j,k}
    \left\{
    y_{i,j}
    -
    \bx_{i,k}^\top \bbeta_{c_{i,k},t_i}
    \right\}^2 .
    \label{eq:objective}
\end{align}
This residual should be interpreted as a working residual for estimating local plaque--cell associations, rather than as a claim that each individual nearby cell independently generates the plaque outcome.
Thus, nearby plaque--cell pairs contribute more strongly to the estimating criterion, whereas distant pairs contribute little or not at all.

{\it In-fill asymptotic justification of the model:} This proximity-weighted construction has a clear population interpretation. Fix a cell
type $c$ and time $t$, and suppose $\boldsymbol{x}^{\star}_{c,t}(\cdot)$ is continuous in
space. As the bandwidth $h_i \to 0$, only cells in a vanishing neighborhood of each
plaque center retain nonzero weight, so the type-$c$ predictors entering the criterion
converge to $\boldsymbol{x}^{\star}_{c,t}(\mathbf{U}_{i,j})$; as the number of sections
$n$ grows, the kernel-weighted normal equations then identify the same coefficient as the
co-located model~\eqref{eq:ideal-model}. In this regime $\boldsymbol{\beta}_{c,t}$ estimates the
marginal co-located regression coefficient of plaque size on local type-$c$ expression,
and the finite-bandwidth estimator used in practice is a smoothed surrogate that trades a
small approximation bias, governed by the spatial variation of
$\boldsymbol{x}^{\star}_{c,t}$ over the kernel support, against the variance reduction
from aggregating nearby cells. We do not pursue a formal rate here; the point is that
``local analogue'' refers to this well-defined target rather than to an estimand created
by the estimator itself.

Next, we specify the kernel function $K_h(\cdot)$. We use a radial kernel that depends
only on the Euclidean distance between a plaque center and a cell location. This choice
treats spatial proximity isotropically, without imposing a preferred direction in the
tissue section, which is appropriate because the plaque-associated cellular organization
documented in Section~\ref{subsec:2:2} is concentric rather than directional. 

Specifically, we use the two-dimensional radial Epanechnikov kernel. This kernel
is compactly supported, so plaque--cell pairs beyond the bandwidth receive zero
weight. This property restricts the estimating criterion to local plaque
neighborhoods and avoids assigning nonzero weights to all possible plaque--cell
pairs. Within the bandwidth, the Epanechnikov kernel decreases smoothly with
distance, so cells closer to a plaque have greater influence than cells near the
boundary of the local neighborhood.

The unscaled two-dimensional radial Epanechnikov kernel is defined as
\[
K(d)
=
\frac{2}{\pi}(1-d^2)_+,
\qquad d\geq 0,
\]
where \(d\) denotes Euclidean distance and \((a)_+=\max(a,0)\). Given a
bandwidth parameter \(h>0\), the scaled kernel is
\[
K_h(d)
=
\frac{1}{h^2}K\left(\frac{d}{h}\right)
=
\frac{2}{\pi h^2}
\left(1-\frac{d^2}{h^2}\right)_+ .
\]
The factor \(h^{-2}\) reflects that the spatial locations are two-dimensional:
increasing the bandwidth expands the kernel neighborhood in both coordinate
directions, so the area scales with \(h^2\). This normalization keeps the total
kernel mass on a comparable scale across different bandwidths.

The factor $h^{-2}$ also has a modeling consequence, not only a normalizing one. Within a
single section a global rescaling of the loss leaves the minimizer unchanged, so the
normalization is immaterial there. It matters across sections: because the sample-specific
bandwidths $h_i$ differ, $h_i^{-2}$ controls the relative contribution of each section to
the pooled criterion, and it also makes the unpenalized losses $\mathrm{Loss}(L)$
comparable across candidate neighborhood sizes in the bandwidth selection. Holding neighborhood size fixed, a section with a smaller
bandwidth (denser cells, shorter median neighbor distance) receives a larger per-pair
weight. This is the intended behavior: the $h_i^{-2}$ scaling places sections on a common
kernel-mass footing so that densely sampled sections do not dominate the fit purely
through cell count, and instead each section contributes according to its local plaque
neighborhoods rather than its raw number of plaque--cell pairs.

The kernel-weighted loss in~\eqref{eq:objective} involves the collection of
cell-type- and time-specific regression coefficients
\[
\{\bbeta_{c,t}\in\mathbb{R}^p:
c=1,\ldots,C,\ t=1,\ldots,T\}.
\]
Equivalently, these coefficients can be arranged into a three-way tensor
\(
\mathcal B \in \mathbb R^{p\times C\times T},
\)
where the first mode indexes genes, the second mode indexes cell types, and the
third mode indexes collection time points. The \((l,c,t)\)-th entry of
\(\mathcal B\) is denoted by \(\beta_{l,c,t}\), representing the regression
effect of gene \(l\) in cell type \(c\) at time point \(t\). 

In the ideal co-located model~\eqref{eq:ideal-model}, \(\beta_{l,c,t}\)
represents the association between plaque size and the expression of gene \(l\)
in cells of type \(c\) at the same spatial location and time point \(t\).
This is the target coefficient we would estimate if plaque outcomes and
cell-level transcriptomic predictors were observed at common locations. In the
observed spatially misaligned setting, such co-located observations are
unavailable. The proposed kernel-weighted criterion therefore estimates
\(\beta_{l,c,t}\) through nearby plaque--cell pairs, with contributions weighted
by spatial proximity. Thus, \(\beta_{l,c,t}\) is interpreted as a local
kernel-weighted analogue of the ideal co-located regression coefficient.

Directly estimating all \(pCT\) coefficients without additional structure can be
statistically unstable, especially when the number of genes is large relative to
the number of independent tissue sections and plaque outcomes. Moreover, an
unstructured model does not borrow information across related cell types or time
points. Thus, we impose a sparse reduced-rank structure
on \(\mathcal B\). This structure serves two purposes: it enables
information sharing across genes, cell types, and time points, and it induces
sparsity in the gene direction to identify genes associated with plaque size.

\paragraph*{Sparse low-rank factorization of the regression coefficient tensor}
To model the coefficient tensor \(\mathcal B\) while borrowing information
across genes, cell types, and time points, we use a reduced-rank tensor
factorization. 
Common tensor factorization methods include the CANDECOMP/PARAFAC (CP) decomposition \citep{carroll_analysis_1970, harshman_foundations_1970, kiers_towards_2000} and the more general Tucker decomposition \citep{tucker_mathematical_1966}. 
In this work, we use a rank-\(R\) CP decomposition of \(\mathcal B\):
\[
\mathcal B
=
\sum_{r=1}^{R}
w_r
\bq^{(1)}_r
\circ
\bq^{(2)}_r
\circ
\bq^{(3)}_r,
\]
or equivalently,
\[
\beta_{l,c,t}
=
\sum_{r=1}^{R}
w_r
q^{(1)}_{l,r}
q^{(2)}_{c,r}
q^{(3)}_{t,r},
\qquad
l=1,\ldots,p,\quad c=1,\ldots,C,\quad t=1,\ldots,T.
\]
Here, \(R\) is the tensor rank, \(w_r>0\) is the weight of the \(r\)-\(th\) rank-one
component, and \(\circ\) denotes the vector outer product. The factor vectors are
\[
\bq^{(1)}_r
=
(q^{(1)}_{1,r},\ldots,q^{(1)}_{p,r})^\top\in\mathbb R^p,
\qquad
\bq^{(2)}_r
=
(q^{(2)}_{1,r},\ldots,q^{(2)}_{C,r})^\top\in\mathbb R^C,
\]
\[
\bq^{(3)}_r
=
(q^{(3)}_{1,r},\ldots,q^{(3)}_{T,r})^\top\in\mathbb R^T
\]
corresponding to the gene, cell-type, and time modes, respectively.

As CP decompositions are invariant to rescaling of the factor vectors, we
impose the normalization constraint on each factor vector:
\[
\|\bq^{(m)}_r\|_2=1,
\qquad
m=1,2,3,\quad r=1,\ldots,R,
\]
where $\|\ba\|_2=\sqrt{\sum_i a_i^2}$ denotes the \(\ell_2\) norm.
This normalization fixes the scale of the factor vectors so that \(w_r\)
captures the overall magnitude of the \(r\)-th rank-one component. It also makes
the sparsity penalty on the gene-mode factors well defined below; without such
normalization, the same tensor could be represented by rescaling one factor up
and another factor down.

The CP representation reduces the dimension of the coefficient tensor and encourages structural sharing across modes. 
Each rank-one component represents a latent pattern consisting of a gene loading vector, a cell-type loading vector, and a time loading vector. 
Therefore, a gene may have different effects across cell types and time points, while the effects are still coupled through shared low-rank components.
Under the CP decomposition, the same rank \(R\) appears across all modes by construction, because \(R\) denotes the number of rank-one components. 
For greater flexibility, Tucker factorization with mode-specific ranks may be considered. 
Here, we use the CP decomposition because it provides a simpler and more directly interpretable sparse tensor regression structure.

To identify important genes, we impose sparsity on the gene-mode factors
\(\{\bq^{(1)}_r\}_{r=1}^R\). 
Specifically, instead of applying an entrywise LASSO penalty directly to all coefficients \(\beta_{l,c,t}\), we apply an \(\ell_1\) penalty to the gene-mode factor vectors. 
The resulting penalized objective is
\begin{align}
    F(\mathcal B)
    =
    \frac{1}{2}
    \sum_{i=1}^n
    \sum_{j=1}^{M_i}
    \sum_{k=1}^{N_i}
    K_{i,j,k}
    \left\{
    y_{i,j}
    -
    \bx_{i,k}^{\top}
    \bbeta_{c_{i,k},t_i}
    \right\}^2
    +
    \frac{\lambda}{Rp}
    \sum_{r=1}^R
    \|\bq_r^{(1)}\|_1,
    \label{eq:objective-penalized}
\end{align}
subject to the CP representation of \(\mathcal B\) and the normalization constraints above. 
Here, \(\lambda\geq 0\) is a tuning parameter controlling the amount of sparsity, and
\[
\|\bq_r^{(1)}\|_1
=
\sum_{l=1}^p |q^{(1)}_{l,r}|
\]
is the \(\ell_1\) norm of the gene-mode factor vector. 
The scaling factor \((Rp)^{-1}\) normalizes the penalty by the number of gene-mode factor
parameters.

This gene-mode penalty provides a natural gene-selection mechanism because the CP
representation propagates sparsity in \(\bq_r^{(1)}\) to the full coefficient
tensor. If
\[
q^{(1)}_{l,r}=0
\qquad
\text{for all } r=1,\ldots,R,
\]
then
\[
\beta_{l,c,t}=0
\qquad
\text{for all } c=1,\ldots,C,\quad t=1,\ldots,T.
\]
Thus, gene \(l\) has no contribution to the regression coefficient tensor. We
therefore regard gene \(l\) as selected if
\[
q^{(1)}_{l,r}\neq 0
\quad
\text{for at least one } r=1,\ldots,R.
\]
In this way, the model selects genes that contribute to at least one latent
component of the coefficient tensor, while allowing the selected genes to have
cell-type- and time-specific effects through the CP structure.

The penalty $\sum_{r}\lVert\mathbf{q}^{(1)}_{r}\rVert_{1}$ acts entrywise within each
component, and a gene is declared selected if it loads on \emph{any} component; the
selected set is therefore the union of the component-wise supports. An alternative is a
row-group penalty $\sum_{l}\lVert\mathbf{q}^{(1)}_{l,\cdot}\rVert_{2}$, which would zero a
gene jointly across all $R$ components and enforce a single common gene support. We prefer
the entrywise form deliberately. Each rank-one component encodes a distinct latent
expression axis with its own cell-type and time loadings, and a gene may participate in
one such axis without participating in the others; the row-group penalty would forbid this
by tying a gene's components together, collapsing biologically distinct programs onto a
shared support. The entrywise penalty instead lets each component recruit its own genes
while still shrinking weak loadings to zero, so the selected set is the union of
component-specific programs rather than a single undifferentiated gene list. This
component-resolved notion of selection is the one we report and interpret in
Section~\ref{sec:5}.

Now, the CP parameterization makes the resulting optimization problem nonconvex, while the sparsity penalty introduces nonsmoothness.
In the next section, we describe the estimation procedure, including initialization, coordinate-wise updates, and selection of the tuning parameters \(\lambda\), \(R\), and \(h_i\).

{\it Thus, different components of the model collectively address the challenges raised in
Section~\ref{sec:1}.} Spatial misalignment between plaque outcomes and cell-level
predictors (challenge~1) is handled by the proximity kernel, which replaces co-located
predictors with distance-weighted neighboring cells. High dimensionality (challenge~2) is managed jointly by the
reduced-rank structure and the gene-mode penalty, the latter also providing interpretable
gene selection. Cell-type-specific effects
(challenge~3, effect heterogeneity across cell types) enter through the cell-type mode of
$\mathcal{B}$, while the shared latent structure of gene effects across cell types and
across the two disease stages (challenges~3 and~4) is captured by the low-rank CP
factorization, which couples otherwise separable cell-type- and time-specific criteria through
common components. The characterization of how neighboring
cells of different types relate to plaque size (challenge~4) is addressed by the kernel-weighted loss. Between-sample
heterogeneity in terms of shift in local cellular composition (challenge~5) is addressed by specifying the sample-specific bandwidths
$h_i$ that absorb differences in cell density and sampling intensity across sections

\subsection{\label{sec:3:1}Estimation of parameters} 

The overall implementation of the estimation procedure is fully automated with data-driven efficient tuning.
We minimize the penalized objective in Eq.~\eqref{eq:objective-penalized} using a block coordinate descent algorithm.
The procedure first collapses the plaque--cell weighted loss into an
equivalent least-squares problem over transformed cell-level rows, then
updates the CP factors one component at a time.
The tuning parameters, including the sparsity parameter \(\lambda\), the tensor rank \(R\), and the bandwidths \(h_i\), are selected as described below.

\paragraph*{Kernel collapse}
For computational convenience, the summation over plaque--cell pairs can be
collapsed before optimization. This transformation is purely algebraic: it does
not change the objective for fixed bandwidths, but rewrites the kernel-weighted
criterion in a rowwise weighted least-squares form.

For fixed tissue section \(i\) and cell \(k\), the fitted value
\(
\bx_{i,k}^{\top}\bbeta_{c_{i,k},t_i}
\)
does not depend on the plaque index \(j\). Therefore, the summation over plaque
locations can be collapsed for each fixed cell. Define the aggregate kernel weight
\[
A_{i,k}
=
\sum_{j=1}^{M_i}K_{i,j,k}.
\]
Let
\(
\mathcal I
=
\{(i,k):A_{i,k}>0\}
\)
denote the set of cells receiving nonzero kernel weight from at least one
plaque. For \((i,k)\in\mathcal I\), define the corresponding kernel-averaged plaque size
around cell \(k\) as
\[
\widetilde y_{i,k}
=
\frac{
\sum_{j=1}^{M_i}K_{i,j,k}y_{i,j}
}{
A_{i,k}
}.
\]
Cells with \(A_{i,k}=0\) do not contribute to the objective and are excluded
from the transformed data.

For fixed \((i,k)\in\mathcal I\), we have
\[
\sum_{j=1}^{M_i}
K_{i,j,k}
\left\{
y_{i,j}
-
\bx_{i,k}^{\top}\bbeta_{c_{i,k},t_i}
\right\}^2
=
C_{i,k}
+
A_{i,k}
\left\{
\widetilde y_{i,k}
-
\bx_{i,k}^{\top}\bbeta_{c_{i,k},t_i}
\right\}^2,
\]
where
\[
C_{i,k}
=
\sum_{j=1}^{M_i}
K_{i,j,k}
\left(
y_{i,j}-\widetilde y_{i,k}
\right)^2
\]
does not depend on the regression coefficients. Thus, up to an additive constant
independent of the model parameters, the kernel-weighted loss is equivalent to
\[
\frac{1}{2}
\sum_{(i,k)\in\mathcal I}
A_{i,k}
\left\{
\widetilde y_{i,k}
-
\bx_{i,k}^{\top}\bbeta_{c_{i,k},t_i}
\right\}^2 .
\]

Equivalently, we define the transformed quantities
\[
y_{i,k}^{\star}
=
A_{i,k}^{1/2}\,\widetilde y_{i,k},
\qquad
\bx_{i,k}^{\star}
=
A_{i,k}^{1/2}\,\bx_{i,k}.
\]
Then, after dropping the parameter-independent constant \(\frac{1}{2}\sum_{(i,k)\in\mathcal I} C_{i,k}\), the penalized objective can be written as
\[
F(\mathcal B)
=
\frac{1}{2}
\sum_{(i,k)\in\mathcal I}
\left\{
y_{i,k}^{\star}
-
{\bx_{i,k}^{\star}}^{\top}\bbeta_{c_{i,k},t_i}
\right\}^2
+
\frac{\lambda}{Rp}
\sum_{r=1}^R
\|\bq_r^{(1)}\|_1,
\]
subject to the CP representation of \(\mathcal B\) and the normalization
constraints on the factor vectors. This least-squares representation is used
only for computation. It should not be interpreted as creating independent
cell-level plaque outcomes, because multiple transformed rows may depend on
overlapping plaque outcomes from the same tissue section.

\paragraph*{Initialization}
As the objective is nonconvex, initialization is important. We initialize
the CP factors using a ridge-regularized estimate of the unstructured
cell-type- and time-specific coefficient tensor. Recall that
\(\mathcal I=\{(i,k):A_{i,k}>0\}\) denotes the transformed rows retained after
the kernel collapse. For each cell type \(c\) and time point \(t\), let
\(
\mathcal I_{c,t}
=
\{(i,k)\in\mathcal I:c_{i,k}=c,\ t_i=t\}.
\)
We define
\[
\widehat \bbeta_{c,t}^{(\mathrm{Ridge})}
\in
\arg\min_{\bbeta\in\mathbb R^p}
\left\{
\sum_{(i,k)\in \mathcal I_{c,t}}
\left(
y_{i,k}^{\star}
-
{\bx_{i,k}^{\star}}^\top\bbeta
\right)^2
+
\lambda_{c,t}\|\bbeta\|_2^2
\right\}.
\]
The vectors \(\widehat \bbeta_{c,t}^{(\mathrm{Ridge})}\in\mathbb R^p\) are
stacked into a three-way tensor
\[
\widehat{\mathcal B}^{(\mathrm{Ridge})}\in\mathbb R^{p\times C\times T},
\qquad
\widehat{\mathcal B}^{(\mathrm{Ridge})}_{:,c,t}
=
\widehat \bbeta_{c,t}^{(\mathrm{Ridge})},
\]
whose \((l,c,t)\)-th entry is the ridge coefficient for gene \(l\), cell type
\(c\), and time point \(t\). A rank-\(R_{\max}\) CP decomposition of
\(\widehat{\mathcal B}^{(\mathrm{Ridge})}\) is then used to initialize
\(\{w_r,\bq_r^{(1)},\bq_r^{(2)},\bq_r^{(3)}\}_{r=1}^{R_{\max}}\).
Here \(R_{\max}\) is the prespecified upper rank used to start the adaptive
CP-rank update; in our implementation, \(R_{\max}=CT\), as discussed later.
The ridge tuning parameter \(\lambda_{c,t}\) is selected by cross-validation.

\paragraph*{Block coordinate updates}
We minimize the transformed penalized objective by block coordinate descent. 
The full parameter set is partitioned into rank-one component blocks
\[
\{w_r,\bq_r^{(1)},\bq_r^{(2)},\bq_r^{(3)}\},
\qquad r=1,\ldots,R.
\]
Within each iteration, we cycle through \(r=1,\ldots,R\) and update one
component at a time while holding the remaining components fixed.

Write
\[
\bbeta_{c,t}
=
\sum_{s=1}^{R}
w_s \bq_s^{(1)} q_{c,s}^{(2)}q_{t,s}^{(3)} .
\]
For component \(r\), we define
\(
\bbeta_{c,t}^{(-r)}
=
\bbeta_{c,t}
-
w_r\bq_r^{(1)}q_{c,r}^{(2)}q_{t,r}^{(3)} .
\)
Using this coefficient, define the component-\(r\) partial residual
\[
e_{i,k}^{(r)}
=
y_{i,k}^{\star}
-
{\bx_{i,k}^{\star}}^\top
\bbeta_{c_{i,k},t_i}^{(-r)},
\qquad (i,k)\in\mathcal I.
\]
This partial residual subtracts the fitted contribution of all components other
than the \(r\)-th component, and is used to update the \(r\)-th rank-one
component.

We first update the gene-mode factor \(\bq_r^{(1)}\). Let
\(x_{i,k,l}^{\star}\) denote the \(l\)-th entry of the transformed predictor
vector \(\bx_{i,k}^{\star}\). For the \(l\)-th coordinate of
\(\bq_r^{(1)}\), define the coordinate-\((l,r)\) partial residual
\[
e_{i,k}^{(l,r)}
=
e_{i,k}^{(r)}
-
w_r q_{c_{i,k},r}^{(2)}q_{t_i,r}^{(3)}
\left\{
{\bx_{i,k}^{\star}}^\top\bq_r^{(1)}
-
x_{i,k,l}^{\star}q_{l,r}^{(1)}
\right\}.
\]
This partial residual subtracts the current contribution of all gene coordinates
except the \(l\)-th coordinate within component \(r\), and is used to update
\(q_{l,r}^{(1)}\).

Let
\(
S_{\tau}(u)=\operatorname{sign}(u)(|u|-\tau)_+
\)
denote the soft-thresholding operator, with \(\tau=\lambda/(Rp)\). The
coordinate-wise update for the gene-mode loading is
\begin{equation}
q_{l,r}^{(1)}
\leftarrow
\frac{
S_{\lambda/(Rp)}
\left[
\displaystyle
\sum_{(i,k)\in\mathcal I}
w_r q_{c_{i,k},r}^{(2)}q_{t_i,r}^{(3)}
x_{i,k,l}^{\star}
e_{i,k}^{(l,r)}
\right]
}{
\displaystyle
\sum_{(i,k)\in\mathcal I}
\left(
w_r q_{c_{i,k},r}^{(2)}q_{t_i,r}^{(3)}
x_{i,k,l}^{\star}
\right)^2
}.
    \label{eq: updating q1}
\end{equation}
If the denominator is zero, we set \(q_{l,r}^{(1)}=0\).

Next, holding the other parameters fixed, the cell-type loading is updated by the closed-form update below. For \(c=1,\ldots,C\),
\begin{equation}
q_{c,r}^{(2)}
\leftarrow
\frac{
\displaystyle
\sum_{(i,k)\in\mathcal I_c}
w_r q_{t_i,r}^{(3)}
{\bx_{i,k}^{\star}}^\top\bq_r^{(1)}
e_{i,k}^{(r)}
}{
\displaystyle
\sum_{(i,k)\in\mathcal I_c}
\left(
w_r q_{t_i,r}^{(3)}
{\bx_{i,k}^{\star}}^\top\bq_r^{(1)}
\right)^2
}.
    \label{eq: updating q2}
\end{equation}
If the denominator is zero, we set \(q_{c,r}^{(2)}=0\).

Similarly, for \(t=1,\ldots,T\), the time loading is updated by
\begin{equation}
q_{t,r}^{(3)}
\leftarrow
\frac{
\displaystyle
\sum_{(i,k)\in\mathcal I_t}
w_r q_{c_{i,k},r}^{(2)}
{\bx_{i,k}^{\star}}^\top\bq_r^{(1)}
e_{i,k}^{(r)}
}{
\displaystyle
\sum_{(i,k)\in\mathcal I_t}
\left(
w_r q_{c_{i,k},r}^{(2)}
{\bx_{i,k}^{\star}}^\top\bq_r^{(1)}
\right)^2
}.
    \label{eq: updating q3}
\end{equation}
If the denominator is zero, we set \(q_{t,r}^{(3)}=0\).

Finally, the component weight \(w_r\geq 0\) is updated by nonnegative least
squares:
\begin{equation}
w_r
\leftarrow
\max\left\{
0,
\frac{
\displaystyle
\sum_{(i,k)\in\mathcal I}
q_{c_{i,k},r}^{(2)}
q_{t_i,r}^{(3)}
{\bx_{i,k}^{\star}}^\top\bq_r^{(1)}
e_{i,k}^{(r)}
}{
\displaystyle
\sum_{(i,k)\in\mathcal I}
\left(
q_{c_{i,k},r}^{(2)}
q_{t_i,r}^{(3)}
{\bx_{i,k}^{\star}}^\top\bq_r^{(1)}
\right)^2
}
\right\}.
    \label{eq: updating w}
\end{equation}
If the denominator is zero, we set \(w_r=0\).

\paragraph*{Computational efficiency and numerical stability}
The kernel collapse reduces the computational burden of the coordinate updates.
A naive implementation of the original objective would evaluate residuals over
all plaque--cell pairs in the triple summation, with per-coordinate cost on the
order of \(O(pN_{\mathrm{all}})\), where
\(N_{\mathrm{all}}=\sum_i M_iN_i\). If the compact support of the kernel is used
to skip zero-weight pairs, this cost becomes \(O(pN_{\mathrm{nz}})\), where
\(N_{\mathrm{nz}}\) is the number of nonzero plaque--cell pairs. After the
kernel collapse, the optimization is instead carried out over
\(N_{\mathrm{row}}=|\mathcal I|\) transformed cell-level rows.

In implementation, we store the current transformed-row residual
\[
\rho_{i,k}
=
y_{i,k}^{\star}
-
{\bx_{i,k}^{\star}}^\top\bbeta_{c_{i,k},t_i},
\qquad (i,k)\in\mathcal I.
\]
When a scalar parameter is updated, these residuals are updated incrementally
using only the change in the fitted contribution of that scalar parameter. For
example, if \(q_{l,r}^{(1)}\) changes by \(\Delta\), then
\[
\rho_{i,k}
\leftarrow
\rho_{i,k}
-
w_rq_{c_{i,k},r}^{(2)}q_{t_i,r}^{(3)}
x_{i,k,l}^{\star}\Delta .
\]
Thus, after the kernel collapse, each scalar coordinate update only requires a pass
over the transformed rows, reducing the cost to \(O(N_{\mathrm{row}})\). The normalization step described below is applied for
numerical stability: it removes arbitrary scale variation among the CP factors and keeps the sparsity penalty on the gene-mode factors well defined.

\paragraph*{Normalization and convergence}
After updating the factors for each rank-one component, we rescale the factor
vectors to have unit \(\ell_2\) norm and absorb the scale into the component
weight \(w_r\). Let
\[
\alpha_{m,r}=\|\bq_r^{(m)}\|_2,
\qquad m=1,2,3.
\]
If \(\alpha_{m,r}>0\) for all \(m=1,2,3\), we set
\[
\bq_r^{(m)}
\leftarrow
\frac{\bq_r^{(m)}}{\alpha_{m,r}},
\qquad m=1,2,3,
\]
and
\[
w_r
\leftarrow
w_r\alpha_{1,r}\alpha_{2,r}\alpha_{3,r}.
\]
This rescaling leaves the fitted coefficient tensor unchanged because the total
rank-one contribution
\[
w_r\bq_r^{(1)}\circ\bq_r^{(2)}\circ\bq_r^{(3)}
\]
is preserved. The normalization removes arbitrary scaling among the CP factors
and keeps the sparsity penalty on the gene-mode factors well defined.

The iterations are stopped when the fitted coefficient tensor becomes stable.
Specifically, we stop when
\begin{equation}
\max_{c,t}
\frac{
\|\bbeta_{c,t}^{(\mathrm{new})}
-
\bbeta_{c,t}^{(\mathrm{old})}\|_{\infty}
}{
\max\{\|\bbeta_{c,t}^{(\mathrm{old})}\|_{\infty},\varepsilon\}
}
<
\delta_{\mathcal B},
    \label{eq: stop rules 1}
\end{equation}
or if the factor updates are sufficiently small,
\begin{equation}
\max_{r,m}
\left\|
\bq_r^{(m,\mathrm{new})}
-
\bq_r^{(m,\mathrm{old})}
\right\|_2
<
\delta_q,
    \label{eq: stop rules 2}
\end{equation}
where \(\varepsilon>0\) is a small stabilizing constant. In our implementation,
we use \(\delta_{\mathcal B}=10^{-6}\) and \(\delta_q=10^{-4}\).

\paragraph*{Adaptive CP-rank update}
To avoid fitting separate models over a large grid of candidate CP ranks, we
embed rank adaptation within the block coordinate descent algorithm. The
algorithm is initialized with a prespecified upper rank \(R_{\max}\). For a
three-way tensor \(\mathcal X\in\mathbb R^{I\times J\times K}\), one general
upper bound on the CP rank is
\[
\operatorname{rank}(\mathcal X)
\leq
\min\{IJ,IK,JK\};
\]
see, for example, \citet{kruskal1989rank} and \citet{kolda_tensor_2009}. For
the coefficient tensor \(\mathcal B\in\mathbb R^{p\times C\times T}\), this
bound becomes
\begin{equation}
\operatorname{rank}(\mathcal B)
\leq
\min\{pC,pT,CT\}.
    \label{eq:rank}
\end{equation}
As \(p\) is typically much larger than \(C\) and \(T\), we use
\(R_{\max}=CT\) as a practical upper rank.

During the initial phase of the optimization, the algorithm updates the active set of rank-one
components by removing numerically negligible components.
Specifically, component \(r\) is removed if
\begin{equation}
w_r=0,
\qquad
\frac{w_r}{\sum_{s=1}^{R}w_s}<\epsilon_R.
    \label{eq:rankdrop}
\end{equation}
In our implementation, we use \(\epsilon_R=10^{-5}\).
We apply this adaptive rank update during
the first \(N_{\mathrm{rank}}\) iterations. Typically, we find that setting \(N_{\mathrm{rank}}=500\) works well.
These conditions remove components with zero or negligible fitted
contribution. In addition, we also drop a rank if 
$\|\bq_r^{(1)}\|_\infty>0.99$. This condition is used as a numerical safeguard against
degenerate components whose gene-mode loading collapses almost entirely onto a
single coordinate during early iterations. After this initial stage, the remaining
components are updated until convergence.

\paragraph*{Selection of the sparsity parameter}
For each candidate bandwidth setting, we fit the model over a decreasing path of
sparsity parameters and select the final \(\lambda\) using a BIC-type criterion. To
construct the path, we start from a sufficiently large value of \(\lambda\) that
yields an all-zero solution and then progressively decrease it,
as in pathwise algorithms for penalized regression
\citep{efron2004least,Friedman2010}. Specifically, we use
\[
\lambda_{k+1}=0.9\lambda_k .
\]
The factor \(0.9\) yields a sufficiently dense grid to capture changes in the fitted model while keeping the computation manageable. In our simulation studies, we find that this choice provides a good balance between accuracy and computational cost.

For each candidate \(\lambda\), we evaluate the BIC-type criterion
\begin{align}
    \mathrm{BIC}(\lambda;L)
    =
    N^*
    \log \left[
    \frac{1}{N^*}
    \sum_{i=1}^n
    \sum_{j=1}^{M_i}
    \sum_{k=1}^{N_i}
    K_{i,j,k}
    \left\{
    y_{i,j}
    -
    \bx_{i,k}^\top
    \widehat\bbeta_{c_{i,k},t_i}
    \right\}^2
    \right]
    +
    \nu\log(N^*),
    \label{eq: lambda selection criterion}
\end{align}
where
\(
N^*
=
\sum_{i=1}^n
\sum_{j=1}^{M_i}
\sum_{k=1}^{N_i}
\mathbf 1\{K_{i,j,k}>0\}
\)
is the number of nonzero plaque--cell pairs, and \(\nu\) is an approximate
degrees of freedom for the fitted sparse CP model. We use
\(
\nu
=
\sum_{r=1}^{\widehat R}
\left(
\|\widehat\bq_r^{(1)}\|_0
+
C
+
T
+
1
\right),
\)
where \(\widehat R\) is the retained rank after the adaptive CP-rank update.
For each fixed bandwidth setting \(L\), the selected sparsity parameter is
\[
\widehat\lambda(L)
=
\arg\min_{\lambda}\mathrm{BIC}(\lambda;L).
\]

\paragraph*{Bandwidth selection}
The kernel bandwidth \(h_i\) is selected through a candidate neighborhood size
\(L\). For plaque \(j\) in sample \(i\), let
\(
D_{i,j}^{(1)} \leq D_{i,j}^{(2)} \leq \cdots \leq D_{i,j}^{(N_i)}
\)
denote the sorted distances from plaque location \(\bU_{i,j}\) to all cell
locations \(\{\bV_{i,k}\}_{k=1}^{N_i}\), that is, the sorted values of
\(\{\|\bU_{i,j}-\bV_{i,k}\|_2:k=1,\ldots,N_i\}\). For a candidate neighborhood
size \(L\), we define the sample-specific bandwidth
\[
H_i(L)=\operatorname{median}_{j=1,\ldots,M_i}D_{i,j}^{(L)}.
\]
Thus, the same neighborhood size \(L\) is used across samples, while the actual
bandwidths \(H_i(L)\) may differ across samples to account for differences in
cell density, as in variable-bandwidth local smoothing
\citep{fan1996study,brunsdon2002geographically}.

For each \(L\), we set \(h_i=H_i(L)\), fit the model over the penalty path, and
select \(\widehat\lambda(L)\) using the BIC-type criterion in
Eq.~\eqref{eq: lambda selection criterion}. We then compute the normalized
unpenalized kernel-weighted loss
\[
\mathrm{Loss}(L)
=
\frac{1}{N^*(L)}
\sum_{i=1}^n
\sum_{k=1}^{N_i}
\sum_{j=1}^{M_i}
K_{H_i(L)}(\|\bU_{i,j}-\bV_{i,k}\|_2)
\left\{
y_{i,j}
-
\bx_{i,k}^\top
\widehat\bbeta_{c_{i,k},t_i}
\right\}^2,
\]
where
\[
N^*(L)
=
\sum_{i=1}^n
\sum_{k=1}^{N_i}
\sum_{j=1}^{M_i}
\mathbf 1
\left[
K_{H_i(L)}(\|\bU_{i,j}-\bV_{i,k}\|_2)>0
\right].
\]
Finally, we select \(\widehat L\) using an automated elbow rule. Specifically, after rescaling both \(L\) and \(\mathrm{Loss}(L)\) to a common range, we choose the value of \(L\) that maximizes the perpendicular distance from the loss curve
\(
{(L,\mathrm{Loss}(L)):L\in\mathcal L}
\)
to the line segment connecting its two endpoints. This criterion identifies the point at which additional increases in \(L\) yield diminishing reductions in the loss, thereby providing a reproducible balance between model fit and computational complexity \citep{salvador2004determining,satopaa2011finding}.

\begin{algorithm}[htbp]
\caption{Comprehensive Implementation Algorithm for the Proposed Method}
    \label{alg:alg}
        \begin{algorithmic}[1]
        \StepHead{Step 1} Select a sequence of $\lambda$ and $L$. For each choice of $(\lambda, L)$ run the following:
        \Require $\{y(\bU_{i,j})\}$, $\{\bx(\bV_{i,k})\}$, $R_{\max}$, $\lambda$, $h_i(L)$;
        construct $(y_{i,k}^{\star},\bx_{i,k}^{\star})$ by the kernel collapse.
        \State (Initialization) Run Ridge regression on the transformed data and get $\hat{\mathcal{B}}^{(\text{Ridge})}$ as described in Sec~\ref{sec:3:1},
        where $\lambda_{\text{Ridge}}$ is selected via cross-validation \citep{Friedman2010}.
        \While{$R_{\max} \ge 1$} \label{ln:init}
             \State $R \gets R_{\max}$
             \State Initialize $w_r$, $\bq^{(1)}_r$, $\bq^{(2)}_r$, and $\bq^{(3)}_r$ by applying CP on $\hat{\mathcal{B}}^{(\text{Ridge})}$ with initial rank $R$.
            \Repeat
                \State Update parameters as equations from Eq.\eqref{eq: updating q1} to Eq.\eqref{eq: updating w}.
                \State $w_r \gets w_r \|\bq^{(1)}_r\|\|\bq^{(2)}_r\|\|\bq^{(3)}_r\|$; $\bq^{(m)}_r \gets \bq^{(m)}_r/\|\bq^{(m)}_r\|$ for $m=1,2,3$.
                \State Update CP-rank $R$ automatically by Eq.\eqref{eq:rankdrop}. 
            \Until{Stopping rules Eq.\eqref{eq: stop rules 1} and Eq.\eqref{eq: stop rules 2} satisfied, or $R=0$}. \label{ln:stop}
            \If{$R=0$}
                \State $R_{\max} \gets R_{\max} - 1$; \textbf{Repeat} Steps 2-9
            \Else
                \State $R_{\mathrm{final}} \gets R$; \textbf{break} \Comment{converged at current rank $R$}
            \EndIf
        \EndWhile
        \noindent
        \Ensure Obtain final estimate as $\mathcal{B}_{\lambda,L}^{(\mathrm{final})} = \sum_{r=1}^{R_{\mathrm{final}}} w_r \bq^{(1)}_r \circ \bq^{(2)}_r \circ \bq^{(3)}_r$ for a given $\lambda$ and $L$.
        
        \StepHead{Step 2} Select the optimal $\hat{\lambda}$ from the pre-specified sequence for each choice of $L$ using the criterion in Eq.~\eqref{eq: lambda selection criterion}.

        \StepHead{Step 3} Select the optimal $\hat{L}$ by applying the elbow rule, and obtain the final estimate $\mathcal{B}^{(\mathrm{final})}_{\hat{\lambda},\hat{L}}$.
    \end{algorithmic}

\end{algorithm}

\section{Simulations}\label{sec:4}

We run two simulations to evaluate the performance of the proposed method in identifying important genes and also estimating their effects. 
Both studies generate realistic single-cell spatial transcriptomics data including cell locations, gene expression, and spatial smoothness by applying established simulators. Subsequently, we generate the outcomes with the spatial misalignment as in our motivating application.

The two settings primarily differ in how the active predictors are set. Also, for the first setting, the SRT data is generated using the \texttt{SRTsim} \citep{zhu2023srtsim} package, and the active predictors are set at random. Thus, their local spatial signal is incidental and often limited. \texttt{SRTsim} is lightweight and easy to use without needing to use any reference real data.
Then the SRT data for the second setting are generated using the \texttt{scDesign3} \citep{song2024scdesign3} package, which can ensure that the fitted data-generating model
preserves predictor-specific marginal distributions, dependence across
predictors, and spatial variation.
Here, the active predictors are also set as the ones with strong local spatial signals. It is motivated by the biological expectation that genes strongly associated with plaque size exhibit spatial organization around plaques.

In both studies, we evaluate variable-selection performance using the true
positive rate and false positive rate over a sequence of thresholds
\(\tau\geq 0\):
\begin{align*}
\operatorname{TPR}(\tau)
&=
\frac{
\sum_{l,c,t}
\mathbf{1}
\left\{
\left|\widehat{\beta}_{l,c,t}\right|\geq\tau,\,
\beta_{l,c,t}\neq0
\right\}
}{
\sum_{l,c,t}
\mathbf{1}
\left\{
\beta_{l,c,t}\neq0
\right\}
},
\\
\operatorname{FPR}(\tau)
&=
\frac{
\sum_{l,c,t}
\mathbf{1}
\left\{
\left|\widehat{\beta}_{l,c,t}\right|\geq\tau,\,
\beta_{l,c,t}=0
\right\}
}{
\sum_{l,c,t}
\mathbf{1}
\left\{
\beta_{l,c,t}=0
\right\}
}.
\end{align*}
Thresholding the estimated coefficients yields a receiver operating
characteristic curve, and the corresponding area under the curve (AUC)
summarizes recovery of the nonzero coefficients. We measure overall
coefficient-estimation accuracy using
\[
\operatorname{MSE}
=
\frac{1}{pCT}
\sum_{l=1}^{p}
\sum_{c=1}^{C}
\sum_{t=1}^{T}
\left(
\widehat{\beta}_{l,c,t}
-
\beta_{l,c,t}
\right)^2.
\]

\vspace{15mm}

\subsection{Simulation setting 1}
\label{sec:sim-app}

\paragraph*{Simulation design}

Following the motivating application, we consider two time points. For each
simulated dataset, we generate two square spatial
transcriptomics samples, one at each time point. For sample
\(i\in\{1,2\}\), let \(\mathcal{S}_i\) denote the set of spatial spots and
let \(t_i=i\) denote the corresponding time stamp. Each sample contains
approximately \(\lvert\mathcal{S}_i\rvert=5000\) spots partitioned into
\(C=3\) cell-type groups by setting the number of groups to
three. The simulator returns a gene-by-spot expression matrix
\(
\mathbf{X}_i
\in
\mathbb{R}^{p\times\lvert\mathcal{S}_i\rvert},
\)
two-dimensional spatial coordinates
\(\{\mathbf{s}_{i,j}:j\in\mathcal{S}_i\}\), and group labels
\(c_{i,j}\in\{1,2,3\}\). We use
\(\mathbf{x}_{i,j}\in\mathbb{R}^{p}\) to denote the expression vector at
spot \(j\), corresponding to the \(j\)-th column of \(\mathbf{X}_i\).

For each sample \(i\), we designate a set of simulated plaque sites
\(\mathcal{P}_i\subset\mathcal{S}_i\), where
\(
M_i
=
\lvert\mathcal{P}_i\rvert
\in
\{50,100,200\}.
\)
These sites are selected to be well separated spatially and approximately
balanced across the three cell-type groups. Specifically, defining
\(
\mathcal{P}_{i,c}
=
\left\{
j\in\mathcal{P}_i:c_{i,j}=c
\right\},
\)
we choose the plaque sites so that \(\lvert\mathcal{P}_{i,c}\rvert\) is
approximately equal across \(c=1,\ldots,C\). The remaining locations,
\(
\mathcal{C}_i
=
\mathcal{S}_i\setminus\mathcal{P}_i,
\)
serve as the observed predictor locations. This construction reproduces the
spatial misalignment in the motivating data: outcomes are observed at plaque
locations, whereas predictor measurements are available only at nearby cell
locations.

We set \(p=50\), designate five predictors as active, and construct the
coefficient tensor using a rank-four CP decomposition. We select five rows of
the predictor-mode factor matrix \(\mathbf{Q}_1\) uniformly at random and
generate their entries independently from \(\mathcal{N}(0,2^2)\); all
remaining rows are set to zero. The entries of the other two factor matrices
are generated independently according to
\(
\mathbf{Q}_2:\ \mathcal{N}(5,2^2),
\mathbf{Q}_3:\ \mathcal{N}(0,0.5^2).
\)
This construction yields coefficient vectors
\(\boldsymbol{\beta}_{c,t}\in\mathbb{R}^{p}\) that vary across cell types and
time points while sharing a sparse, low-rank structure.

For each designated plaque site \(j\in\mathcal{P}_i\), we generate a
synthetic plaque-level outcome as
\[
y_{i,j}
=
\mathbf{x}_{i,j}^{\top}
\boldsymbol{\beta}_{c_{i,j},t_i}
+
e_{i,j}.
\]
Let \(\mathbf{e}_i=(e_{i,j})_{j\in\mathcal{P}_i}\). To incorporate residual
spatial dependence, we generate
\[
\mathbf{e}_i
\sim
\mathcal{N}
\left(
\mathbf{0},
\sigma_e^2\mathbf{K}_i
\right),
\qquad
\sigma_e^2
\in
\{1,5,10,100,200\},
\]
where
\[
K_i(j,j')
=
\exp
\left\{
-\frac{d_i(j,j')}{\widehat{\phi}}
\right\},
\]
and \(d_i(j,j')\) is the Euclidean distance between plaque sites \(j\) and
\(j'\). We estimate \(\widehat{\phi}\) once from the real plaque data by
fitting an exponential variogram with the \texttt{variog} and
\texttt{variofit} functions in the \texttt{geoR} package \citep{R-geoR}, and
then hold it fixed across all simulation settings and replicates.

Although the outcome \(y_{i,j}\) is generated using the expression vector
\(\mathbf{x}_{i,j}\) at the same location, this vector is treated as
unobserved during estimation. Specifically, we remove the expression
measurements
\(
\left\{
\mathbf{x}_{i,j}:j\in\mathcal{P}_i
\right\}
\)
from the predictor data and model the outcomes
\(
\left\{
y_{i,j}:j\in\mathcal{P}_i
\right\}
\)
using only the expression measurements at non-plaque locations,
\(
\left\{
\mathbf{x}_{i,k}:k\in\mathcal{C}_i
\right\}.
\)
Thus, the estimator must recover the coefficient structure by borrowing
information from predictor measurements at neighboring spatial locations.

Because the active predictors are selected independently of their spatial structure, some may carry weak local spatial signal: once their values at the outcome locations are withheld, neighboring measurements carry little information about them. This is attributed to the simulation design. 

\paragraph*{Estimation and comparison}

We apply the estimation procedure described in Algorithm~\ref{alg:alg}. For
each candidate bandwidth, the procedure estimates a sparse, low-rank
coefficient tensor over a sequence of penalty parameters. The penalty
parameter is selected using the BIC criterion in
Equation~\eqref{eq: lambda selection criterion}, and the bandwidth is selected
using the elbow rule described in Section~\ref{sec:method}.

Following the rank characterization in Equation~\eqref{eq:rank}, we set the
maximum candidate rank to \(R_{\max}=6\). We index the bandwidth by the number
of neighboring predictor locations and consider
\(
L
\in
\{5,10,12,15,20,25,30,35,40,45,50,70\}.
\)
The penalty sequence begins at
\(
\lambda_{\max}
=
10^{10}\times 0.9^{-30}
\)
and decreases geometrically. The rank, penalty parameter, and bandwidth are
therefore selected entirely from the observed simulated data.

We compare the proposed estimator with a nearest-neighbor-paired LASSO baseline. Specifically, each
simulated plaque-level outcome \(y_{i,j}\) is paired with the expression vector
at its nearest non-plaque location,
\(
\mathbf{x}_{i,k^\ast},\) where \(
k^\ast
=
\underset{k\in\mathcal{C}_i}{\arg\min}\,
\left\|
\mathbf{s}_{i,j}
-
\mathbf{s}_{i,k}
\right\|_2.
\)
We then analyze the resulting paired observations using an
\(\ell_1\)-regularized linear regression model. 

We then vary $M_i$ and $\sigma_e^2$ for different simulation settings. For each combination of \(M_i\) and \(\sigma_e^2\), we report the MSE and AUC
averaged over 30 independent simulation replicates.

\paragraph*{Results}
Table~\ref{tab:simulation-results-1} summarizes the results of the study. Across every combination of the number of outcome
locations and the error variance, the proposed method yields a lower
coefficient MSE and a higher AUC than the paired LASSO. The paired LASSO shows little discriminatory ability and frequently produces coefficient estimates close to zero, whereas the proposed method distinguishes active from inactive
coefficients across the full range of settings. The advantage is most
pronounced when more outcome locations are available and the error variance is
moderate.

The absolute AUC values are modest, particularly when the number of outcome
locations is small or the error variance is large. This reflects the
difficulty of the design rather than a shortcoming of the estimator. Because
the active predictors are selected without regard to their local spatial
signal, neighboring observations may contain limited information about some
unobserved active predictors at the outcome locations; when a predictor is
spatially unstructured, its value at an outcome location is simply not encoded
by nearby measurements, so no estimator can recover it.
The simulation setting in Section~\ref{sec:sim-controlled} examines the estimation when the active predictors are set based on their spatial signal strength.

\begin{table}[!htbp]
\centering
\footnotesize
\setlength{\tabcolsep}{4pt}
\begin{tabular}{@{}l l c c c c@{}}
\toprule
&
&
\multicolumn{2}{c}{MSE}
&
\multicolumn{2}{c}{AUC}
\\
\cmidrule(lr){3-4}
\cmidrule(lr){5-6}
\(M_i\)
&
\(\sigma_e^2\)
&
Proposed
&
LASSO
&
Proposed
&
LASSO
\\
\midrule
50  & 1   & 6.333 & 7.086 & 0.574 & 0.485 \\
50  & 5   & 6.367 & 7.014 & 0.563 & 0.485 \\
50  & 10  & 6.392 & 6.985 & 0.544 & 0.484 \\
50  & 100 & 6.891 & 8.029 & 0.495 & 0.501 \\
50  & 200 & 7.359 & 9.169 & 0.510 & 0.507 \\
\addlinespace
100 & 1   & 6.256 & 7.455 & 0.647 & 0.514 \\
100 & 5   & 6.253 & 7.422 & 0.629 & 0.524 \\
100 & 10  & 6.275 & 7.380 & 0.650 & 0.525 \\
100 & 100 & 6.403 & 8.330 & 0.565 & 0.539 \\
100 & 200 & 6.586 & 9.256 & 0.568 & 0.546 \\
\addlinespace
200 & 1   & 6.278 & 6.706 & 0.626 & 0.484 \\
200 & 5   & 6.235 & 6.717 & 0.607 & 0.489 \\
200 & 10  & 6.273 & 6.705 & 0.606 & 0.494 \\
200 & 100 & 6.319 & 6.815 & 0.565 & 0.501 \\
200 & 200 & 6.486 & 8.356 & 0.568 & 0.498 \\
\bottomrule
\end{tabular}
\caption{Simulation setting 1 performance over 30 replicates for
different numbers of simulated plaque sites \(M_i\) and marginal error
variances \(\sigma_e^2\). ``Proposed'' denotes the kernel-weighted sparse
reduced-rank estimator, and ``LASSO'' denotes the baseline obtained by matching
each plaque-level outcome to its nearest non-plaque spot. MSE is computed over
the coefficient tensor, and AUC summarizes variable-selection performance
based on thresholded coefficient estimates.}
\label{tab:simulation-results-1}
\end{table}

\subsection{Simulation setting 2}
\label{sec:sim-controlled}

\paragraph*{Simulation design}

Here, we retain the data-generating pipeline of Setting~1 but change how the active predictors are set. Specifically, we designate the genes with the strongest local spatial signal as active. We then vary the noise variance across different simulation settings.

For this simulation, we set \(p=200\), \(C=4\), and \(T=2\), giving a coefficient tensor
\(
\mathcal{B}
\in
\mathbb{R}^{200\times 4\times 2}.
\)
As in Simulation setting~1, the true coefficient tensor has CP rank four and
is sparse in the predictor mode. Here, however, we designate 20 predictors as
active by selecting those with the largest latent spatial amplitudes, measured
by the spatial variances of their fitted log-rate surfaces. We generate the
corresponding rows of \(\mathbf{Q}_1\) independently from
\(\mathcal{N}(0,2^2)\) and set the remaining rows to zero. The entries of the
other factor matrices are generated independently according to
\(
\mathbf{Q}_2:\ \mathcal{N}(5,1^2),
\mathbf{Q}_3:\ \mathcal{N}(0,0.5^2).
\)
The columns of the factor matrices are normalized, with their scales absorbed
into the corresponding CP weights. This construction produces a sparse,
rank-four coefficient tensor whose active predictors have appreciable spatial
structure.

We generate the complete predictor matrix and true coefficient tensor once
and hold them fixed throughout the study. We select \(M=200\) outcome
locations from areas with high local predictor density, approximately
balancing them across the \(CT=8\) coefficient slices. For outcome location
\(j\), associated with coefficient slice \((c,t)\), we generate
\[
y_j
=
\mathbf{x}_j^{\top}
\boldsymbol{\beta}_{c,t}
+
e_j.
\]
As in Setting 1, \(\mathbf{x}_j\) is used only to
generate the response and is removed before model fitting. The proposed
estimator therefore relies exclusively on predictor measurements observed at
neighboring, spatially distinct locations.

We again generate the errors from the spatially correlated Gaussian distribution with
\(
K(j,j')
=
\exp
\left\{
-\frac{d(j,j')}{\phi}
\right\},\) where \(
\phi
=
200\,\mu\mathrm{m},
\)
and \(d(j,j')\) is the Euclidean distance between outcome locations \(j\) and
\(j'\).
Here \(\phi\) is fixed at \(200\,\mu\mathrm{m}\) rather than estimated from
data as in Setting~1, so that the spatial error scale is held constant across this study.
For each value of \(\sigma_e^2\), we generate 50 independent error
realizations. 

We use the same general estimation and tuning procedure as in Simulation
setting~1. The proposed estimator is again compared with a nearest-neighbor
LASSO baseline. Within each coefficient slice, each outcome location is matched to its nearest observed predictor location, and a separate LASSO model is fitted. 
For each setting, we report the median MSE and AUC over the 50 independent error realizations.

\paragraph*{Results}

Table~\ref{tab:simulation-results-2} summarizes the results of this study. When the active predictors have appreciable spatial
structure, the proposed method achieves an AUC of \(0.808\) for
\(\sigma_e^2\leq 10\), compared with approximately \(0.50\) for the LASSO
baseline. As the error variance increases to 100 and 200, the AUC decreases
gradually to \(0.788\) and \(0.759\), respectively, while the coefficient MSE
increases only slightly.

These results show that the proposed method efficiently recovers the true coefficient structure. They also
show the expected deterioration in variable-selection performance as outcome noise increases. In contrast, the slice-specific LASSO provides essentially
random discrimination between active and inactive coefficients across all
investigated noise levels.

\begin{table}[!htbp]
\centering
\small
\begin{tabular}{crrrr}
\toprule
&
\multicolumn{2}{c}{MSE}
&
\multicolumn{2}{c}{AUC}
\\
\cmidrule(lr){2-3}
\cmidrule(lr){4-5}
\(\sigma_e^2\)
&
Proposed
&
LASSO
&
Proposed
&
LASSO
\\
\midrule
1   & 8.02 & 13.90 & 0.808 & 0.495 \\
5   & 8.03 & 13.95 & 0.808 & 0.495 \\
10  & 8.04 & 13.35 & 0.808 & 0.496 \\
100 & 8.10 & 14.69 & 0.788 & 0.502 \\
200 & 8.27 & 17.27 & 0.759 & 0.504 \\
\bottomrule
\end{tabular}
\caption{Simulation setting 2 performance under increasing
outcome noise, with the number of outcome locations fixed at \(M=200\).
Entries are medians over 50 independent error realizations. Here,
\(\sigma_e^2\) denotes the marginal error variance, MSE is computed over the
complete coefficient tensor, and AUC summarizes recovery of the nonzero
coefficients.}
\label{tab:simulation-results-2}
\end{table}

In summary, the two studies illustrate performance of the proposed
method under two data generation settings. The first study assesses performance when active
predictors are selected without regard to their local spatial signal. Then the second study sets the active set as the most spatially varying genes, which is more realistic from our application's point of view and further evaluates the method's sensitivity to outcome
noise.

\section{\label{sec:5}STARmap PLUS Alzheimer’s Disease Application}

In this section, we perform the integrative spatial transcriptomics analysis to examine the relationship between cell-level gene-expression profiles and Amyloid-$\beta$ (A$\beta$) plaque size in Alzheimer's disease (AD), focusing on two disease stages, 8 and 13 months, and three specific cell types: Astrocytes, Oligodendrocyte-lineage cells, and Microglia. 
We analyze four SRT tissue sections ($i = 1,2,3,4$), with two sections at each disease stage. Specifically, sections $i=1,2$ correspond to the two 8-month replicates, and sections $i=3,4$ correspond to the two 13-month replicates.
The cell counts are $(N_1,N_2,N_3,N_4) = (8{,}186,\ 8{,}202,\ 10{,}372,\ 9{,}634)$, and plaque counts are $(M_1,M_2,M_3,M_4) = (88,\ 99,\ 192,\ 136)$.
We use the normalized expression provided by \cite{zeng_integrative_2023} to place all genes on a common scale across sections and disease stages. 

We set the $R_{\text{max}}$ at $6$, and $\lambda_{\text{max}} = 10^5$. For bandwidth selection, we consider the grid $L\in {1,\dots,20,25,30}$ neighborhood size and set the sample-specific bandwidth as $h_i=H_i(L)$, where $H_i(L)$ is defined in Section~\ref{sec:3:1}. The bandwidth and penalty parameters are then selected using the procedure described in Algorithm \ref{alg:alg}.

Prior to fitting the plaque-size regressions, we apply an expression filter independent of the plaque-size outcome to exclude transcripts with insufficient detection for stable estimation. Specifically, within each of the three target cell-type groups and for each tissue sample collected at 8 and 13 months, we retain genes detected (nonzero) in $>20\%$ of near-plaque cells (within $150 \mu$m of the plaque center). We then augment the retained set with (i) 64 marker genes from \citet{zeng_integrative_2023} and (ii) plaque-induced genes (PIGs) from \citet{chen_spatial_2020} (32 of the 57 PIGs are present in our dataset).  This procedure yields a final gene panel of (p=182) transcripts for modeling the plaque-size outcome.

In the application below, the coefficient tensor is therefore \(\mathcal{B}\in\mathbb{R}^{182\times 3\times 2}\), with a CP structure relating plaque size to local, cell-type-resolved expression across disease stages. 
The final rank is $R=4$, with penalty $\lambda = 10^{5}(0.9)^{21} \approx 1.094\times 10^{4}$. 
The elbow rule selects $L=5$. Accordingly, we set the sample-specific bandwidth for the kernel function as $h_i = H_i(5), i=1,2,3,4$, where $H_i(L)$ denotes the median distance to the $L$-$th$ nearest neighboring cell in sample $i$ (defined in Sec~\ref{sec:3:1}). This yields $h= (41.77,45.42,36.89,37.76) \mu$m.

Here, \(\widehat{\beta}_{l,c,t}\) is interpreted as a local kernel-weighted
analogue of the ideal co-located regression coefficient. It summarizes the
association between plaque size and the normalized expression of
gene \(l\) of cell-type \(c\) at disease stage \(t\) conditioned on other genes of the same cell-type. A positive coefficient
indicates that higher expression of gene \(l\) in proximal cells of type \(c\)
is associated with larger plaque radius under the fitted kernel-weighted
model, whereas a negative coefficient indicates association with smaller plaque
radius. This interpretation is associational and should not be read as implying
that any single nearby cell independently determines plaque size.

For each cell type $c$ and disease stage $t$, we summarize the overall magnitude of the estimated plaque-size association by $A_{c,t}=\|\beta_{\cdot c t}\|_2,$ the $\ell_2$ norm of $p$ gene-specific coefficients. We summarize the averaged signed direction by $\bar \beta_{c,t} = p^{-1} \sum_{l=1}^p \beta_{l,c,t},$ i.e., the mean gene effect across all genes. 
At 8 months, estimated association magnitudes $A_{c,t}$ is modest (Astrocyte: 2.30; Oligodendrocyte: 6.20; Microglia: 3.04), but it increases sharply by 13 months (Astrocyte: 19.40; Oligodendrocyte: 39.74; Microglia: 21.81), indicating substantially stronger plaque--expression associations over time. 
The signed means $\bar \beta_{c,t}$ are near zero at 8 months (Astrocyte: $-0.0049$, Oligodendrocyte $-0.0018$), while Microglia shows a small positive mean ($0.0285$), meaning that higher microglial gene expression tends to be associated with larger plaques on average at 8 months. 
By 13 months, the average effects are predominantly positive for Oligodendrocyte (0.0379) and especially Microglia (0.1536), while Astrocyte remains near zero (0.0017). Overall, these results point to strong, increasing plaque-proximal gene-expression effects in oligodendrocyte-lineage cells and microglia over time, with microglia showing the clearest positive directional signal at 13 months.

Following the CP structure, we pursue a component-level interpretation of the results here.
The CP decomposition represents the estimated coefficient tensor using four rank-one latent components, each combining a gene-loading vector, a cell-type-loading vector, and a disease-stage-loading vector (Table~\ref{tab:cp_components}).
The first three components have substantially larger weights than the fourth. 
For component $r$ with factors $\bq_{r}^{(1)}$ (genes), $\bq_r^{(2)}$ (cell types), $\bq_r^{(3)}$ (times) and weight $w_r$, we orient it so that its top-loading genes have positive entries in the gene factor, and then summarize its overall direction using the signed-average product
\begin{equation*}
    \text{NetDir}_r = \left(\frac{\sum_l q_{l,r}^{(1)}}{\sum_l |q_{l,r}^{(1)}|}\right) \left(\frac{\sum_c q_{c,r}^{(2)}}{\sum_c |q_{c,r}^{(2)}|}\right) \left(\frac{\sum_t q_{t,r}^{(3)}}{\sum_t |q_{t,r}^{(3)}|}\right) \in [-1,1].
\end{equation*}
We refer to 13 months as the late disease stage.
As shown in Table~\ref{tab:cp_components}, each component is dominated by the late time point, with the 13-month time loading having magnitude close to 1, indicating that the estimated associations concentrate their mass at 13 months (with direction determined by the joint sign across modes; under the chosen orientation).
Among them, Component 3 shows a clearly positive direction (Net direction $\approx +0.47$) with largest loadings on Oligodendrocytes and Microglia at 13 months; top genes include Trem2, Tmsb4x, Tyrobp, and Plp1, indicating a late microglia--oligodendrocyte component-derived signature in which higher expression is associated with larger plaques.
In contrast, Components 1 and 2 carry negative temporal loadings at 13 months and split cell-type emphasis--one tilting toward Oligodendrocytes (Oligodendrocyte ↑, Microglia ↓) and the other toward Microglia (Microglia ↑, Astrocyte ↓). Despite substantial overlap in top genes (e.g., Trem2, C1qa, Aplp1), their opposite cell-type signs imply opposing directions of association across components at the same time point, consistent with heterogeneity between microglia-enriched expression patterns and oligodendrocyte/myelin related pathways. 
Finally, Component 4 is uniformly negative (Net direction $\approx -1$), dominated by Oligodendrocyte and Microglia loadings at 13 months with genes such as Tmsb4x, Plp1, C1qa, and Cst3, suggesting a late-stage component pattern in which higher expression accompanies smaller plaques. 
Thus, in summary, the estimated associations concentrate at 13 months and along a Microglia--Oligodendrocyte axis, but with mixed directions across components: Component 3 aligns with larger plaques, whereas Components 1, 2, and 4 align with smaller plaques, suggesting distinct plaque-proximal expression patterns.

The CP-component-based results provide a module-level summary that is broadly consistent with prior biological knowledge. Components involving microglial activation and complement-related genes are consistent with plaque-associated microglial responses reported in AD \citep{hong2016complement,KerenShaul2017Cell,Krasemann2017Immunity}, whereas components involving oligodendrocyte and myelin-related genes are consistent with reported white-matter and myelin alterations in AD \citep{mckenzie2017multiscale,nasrabady2018white}. These summaries capture known biological signals through a small number of signed multigene components localized by cell type and disease stage, providing a component-level perspective that complements the gene-level coefficient analyses discussed next.

\begin{table}[!t]
\centering
\footnotesize
\begin{tabularx}{\textwidth}{@{} c c r >{\hsize=10.8\hsize}Y >{\hsize=6.50\hsize}Y >{\hsize=6.05\hsize}Y @{}}
\toprule
Component,$r$ & Weight, $w_r$ & NetDir & Top cells & Top times & Top genes \\
\midrule
$1$ & $98.27$ & $-0.1319$
& \makecell[tl]{Oligodendrocyte ($+0.75$),\\ Microglia ($-0.51$)}
& \makecell[tl]{13 mo ($-0.99$),\\ 8 mo ($-0.15$)}
& Trem2($+0.91$), Tyrobp($-0.24$), C1qa($-0.19$), Aplp1($+0.13$), Mbp($+0.08$) \\
$2$ & $86.34$ & $-0.0287$
& \makecell[tl]{Microglia ($+0.83$),\\ Astrocyte ($-0.55$)}
& \makecell[tl]{13 mo ($-0.99$),\\ 8 mo ($-0.13$)}
& Trem2($+0.92$), C1qa($-0.20$), Aspa($-0.15$), Aplp1($+0.13$), Tyrobp($-0.11$) \\
$3$ & $66.13$ & $+0.4662$
& \makecell[tl]{Oligodendrocyte ($+0.93$),\\ Microglia ($+0.35$)}
& \makecell[tl]{13 mo ($+0.99$),\\ 8 mo ($+0.12$)}
& Trem2($+0.85$), Tmsb4x($+0.31$), Tyrobp($+0.22$), Plp1($+0.17$), Aplp1($+0.15$) \\
$4$ & $19.49$ & $-1.0000$
& \makecell[tl]{Oligodendrocyte ($-0.85$),\\ Microglia ($-0.52$)}
& \makecell[tl]{13 mo ($+1.00$),\\ 8 mo ($+0.09$)}
& Tmsb4x($+0.85$), Plp1($+0.40$), C1qa($+0.29$), Cst3($+0.19$) \\
\bottomrule
\end{tabularx}
\caption{CP component summary for the estimated coefficient tensor. For each
component \(r\), ``Top cells/times/genes'' are entries with the largest-magnitude
loadings in the corresponding mode. Component signs are oriented so that the
top-loading genes have positive entries. As individual CP factor signs are
not separately identifiable, the direction of association is interpreted through
the reconstructed coefficient contribution; positive values indicate association
with larger plaques and negative values indicate association with smaller
plaques.}
\label{tab:cp_components}
\end{table}

\noindent\textbf{Gene-level examples consistent with prior literature (associational).}
To summarize temporal changes in estimated regression coefficients (associational gene effects), we write $a\!\to\!b$ for gene $l$ in cell type $c$, where $\hat\beta_{l,c,8}=a$ (8 months) and $\hat\beta_{l,c,13}=b$ (13 months).
\underline{\textit{Astrocytes:}} transport-related markers show increasingly negative associations with plaque size (Aqp4 $-0.27\!\to\!-1.85$, Slc13a3 $-0.26\!\to\!-1.67$), while alarmin/stress signals become more positively associated (Il33 $+0.27\!\to\!+1.99$, S100a6 $+0.28\!\to\!+1.82$), consistent with reactive astrocyte frameworks and IL-33--microglia crosstalk \citep{Escartin2021NatNeuro,Vainchtein2018Science}.
\underline{\textit{Oligodendrocyte lineage:}} larger plaques are associated with stronger precursor-like signal (Gpr17 $+0.29\!\to\!+2.39$) and reduced mature oligodendrocyte identity (Sox10 $-0.16\!\to\!-0.50$, Aspa $\approx\!-0.20$). These patterns are consistent with reported links between oligodendrocyte function and iron/transferrin biology \citep{Todorich2009Glia}, and OPC--vasculature coupling provides spatial context \citep{Tsai2016Science}.
\underline{\textit{Microglia:}} an early Trem2-associated effect attenuates and reverses (Trem2 $+0.64\!\to\!-0.37$), while a later Apoe/Tyrobp-centered signature becomes strongly positively associated with plaque size (Apoe $+0.21\!\to\!+1.78$; Tyrobp $-0.11\!\to\!+1.20$), mirroring reported disease-associated microglia (DAM)-like shifts and the TREM2--APOE pathway \citep{KerenShaul2017Cell,Krasemann2017Immunity,Deczkowska2018Cell}.

\noindent\textbf{Novel or less-established associations requiring validation (hypothesis-generating).}
Some additional genes show large late-stage coefficients but have less direct support in the plaque-size regression context; these findings should therefore be viewed as hypothesis-generating.
\ul{\textit{Oligodendrocyte “logistics”:}} strong, late positive coefficients for resource intake and protein production (Trf $+0.53\!\to\!+3.53$, Pabpc1 $+0.50\!\to\!+3.55$, Caskin1 $+0.52\!\to\!+3.95$) suggest a resource-mobilizing oligodendrocyte niche as plaques enlarge; in contrast, Flt1 is strongly negative ($-0.57\!\to\!-4.02$), pointing to a vascular/adhesion signature aligned with smaller plaques.
\ul{\textit{Microglial cytoskeleton and lysosome:}} Rhoc shows a robust negative association ($-0.52\!\to\!-3.95$), consistent with more contractile morphology near smaller plaques, while cathepsins diverge (Ctsl $+0.20\!\to\!+1.04$ vs Ctss $-0.11\!\to\!-0.93$), indicating targeted lysosomal remodeling rather than uniform activation.
\underline{\textit{Astrocytes:}} persistence and amplification of transport↓/alarmin↑ patterns across time underscore a simple, testable hypothesis: supporting basic astrocytic transport may align with smaller plaques.

\section{Conclusion}
\label{sec:conc}
In this paper, we introduce a novel kernel-weighted method to model how single-cell spatial transcriptomic expression patterns relate to plaque size. 
The proposed framework is designed to capture both cell-type-specific and disease-stage-specific gene effects, thereby accommodating heterogeneity across genes, cell populations, and sample collection times. In addition, the method is fully automated through data-driven specification of the tuning parameters, such as rank, penalty, and bandwidth.
We further adopt a low-rank model for the regression coefficients, which naturally captures the shared structure and inherent dependencies among genes, cell types, and time points.

Our analysis reveals several plaque-size-associated expression patterns that are strongest at the 13-month data, mainly in microglia (brain immune cells) and oligodendrocytes (myelin-related cells), indicating that plaque-size associations reflect multiple transcriptional signatures acting in parallel rather than a single dominant signature. 
At the late disease stage, the microglial pattern is characterized by lipid-handling and complement-related genes, consistent with a plaque-proximal activation signature \citep{hong2016complement, KerenShaul2017Cell, Krasemann2017Immunity}; oligodendrocyte-lineage associations suggest a shift toward less mature (precursor-like) signatures, together with increased iron-homeostasis and protein-synthesis transcripts; 
and astrocyte associations are weaker overall and include transport-related genes with negative plaque-size associations.
These findings are associational and should be viewed as hypothesis-generating, motivating future experimental studies of how plaque-proximal glial programs relate to plaque progression.

Future work will extend the framework to model how plaque-associated gene-expression patterns vary with plaque burden, disease stage, and anatomical region.
This will let us map which genes and multigene expression patterns rise or fall as plaques enlarge, compare patterns across different disease stages and brain regions. These insights will have important translational implications for understanding disease mechanisms and guiding hypothesis-driven experimental studies. Methodologically, there are opportunities to consider more flexible Tucker factorization for $\bbeta$, and also other types of effect characterizations beyond linearity.

\paragraph*{Data availability}

We use publicly available data from \cite{zeng_integrative_2023} for our analysis.

\section*{Acknowledgments}
The authors would like to thank Dr. Dongyuan Wu for helpful discussions and suggestions regarding the data. During the preparation of this work, the authors used ChatGPT to assist with writing. All content was subsequently reviewed and edited by the authors, who take full responsibility for the publication.

\bibliographystyle{abbrvnat}
\bibliography{bibliography}

@article{Deczkowska2018Cell,
  title   = {Disease-Associated Microglia: A Universal Immune Sensor of Neurodegeneration},
  author  = {Deczkowska, Aleksandra and Keren-Shaul, Hadas and Weiner, Assaf and Colonna, Marco and Schwartz, Michal and Amit, Ido},
  journal = {Cell},
  year    = {2018},
  volume  = {173},
  number  = {5},
  pages   = {1073--1081},
  doi     = {10.1016/j.cell.2018.05.003}
}

@article{Escartin2021NatNeuro,
  title   = {Reactive astrocyte nomenclature, definitions, and future directions},
  author  = {Escartin, Carmen and Galea, Elena and Lakatos, Andras and O'Callaghan, James P. and Petzold, Axel and Serrano-Pozo, Alberto and Steinh{\"a}user, Christian and Volterra, Andrea and Carmignoto, Giorgio and Agarwal, Arpita and others},
  journal = {Nature Neuroscience},
  year    = {2021},
  volume  = {24},
  pages   = {312--325},
  doi     = {10.1038/s41593-020-00783-4}
}

@Article{Friedman2010,
  author  = {Jerome Friedman and Trevor Hastie and Robert Tibshirani},
  title   = {Regularization Paths for Generalized Linear Models via Coordinate Descent},
  journal = {Journal of Statistical Software},
  year    = {2010},
  volume  = {33},
  number  = {1},
  pages   = {1--22},
  doi     = {10.18637/jss.v033.i01}
}

@article{KerenShaul2017Cell,
  title   = {A Unique Microglia Type Associated with Restricting Development of Alzheimer’s Disease},
  author  = {Keren-Shaul, Hadas and Spinrad, Arik and Weiner, Assaf and Matcovitch-Natan, Orit and Dvir-Szternfeld, Ravid and Ulland, Tyler K. and David, Eyal and Baruch, Kuti and Lara-Astaiso, Diana and Toth, Balint and others},
  journal = {Cell},
  year    = {2017},
  volume  = {169},
  number  = {7},
  pages   = {1276--1290.e17},
  doi     = {10.1016/j.cell.2017.05.018}
}

@article{Krasemann2017Immunity,
  title   = {The TREM2-APOE Pathway Drives the Transcriptional Phenotype of Dysfunctional Microglia in Neurodegenerative Diseases},
  author  = {Krasemann, Susanne and Madore, Camille and Cialic, Rachel and Baufeld, Christian and Calcagno, Nicole and El Fatimy, Rachid and Beckers, Liliane and O'Loughlin, Eoin and Xu, Yungui and Fanek, Zachariah and others},
  journal = {Immunity},
  year    = {2017},
  volume  = {47},
  number  = {4},
  pages   = {566--581},
  doi     = {10.1016/j.immuni.2017.08.008}
}

@article{Todorich2009Glia,
  title   = {Oligodendrocytes and myelination: the role of iron},
  author  = {Todorich, Barbara and Pasquini, Juliana M. and Garcia, Carlos I. and Paez, Pablo M. and Connor, James R.},
  journal = {Glia},
  year    = {2009},
  volume  = {57},
  number  = {5},
  pages   = {467--478},
  doi     = {10.1002/glia.20784}
}

@article{Tsai2016Science,
  title   = {Oligodendrocyte precursors migrate along vasculature in the developing nervous system},
  author  = {Tsai, Hsiang-Hsuan and Niu, Jing and Munji, Rami and Davalos, Dimitrios and Chang, Jason and Zhang, Hongmei and Tien, Ann C. and Kuo, Christopher J. and Chan, Jonah R. and Daneman, Richard and Fancy, Sheep P. J.},
  journal = {Science},
  year    = {2016},
  volume  = {351},
  number  = {6271},
  pages   = {379--384},
  doi     = {10.1126/science.aad3839}
}

@article{Vainchtein2018Science,
  title   = {Astrocyte-derived interleukin-33 promotes microglial synapse engulfment and neural circuit development},
  author  = {Vainchtein, Ilia D. and Chin, Gregory and Cho, Frances S. and Kelley, Kevin W. and Miller, John G. and Chien, Emily C. and Liddelow, Shane A. and Nguyen, Peter T. and Nakao-Inoue, Hiromi and Dorman, Lindsey C. and others},
  journal = {Science},
  year    = {2018},
  volume  = {359},
  number  = {6381},
  pages   = {1269--1273},
  doi     = {10.1126/science.aal3589}
}

@article{alexeeff2016spatial,
  title={Spatial measurement error and correction by spatial SIMEX in linear regression models when using predicted air pollution exposures},
  author={Alexeeff, Stacey E and Carroll, Raymond J and Coull, Brent},
  journal={Biostatistics},
  volume={17},
  number={2},
  pages={377--389},
  year={2016},
  publisher={Oxford University Press}
}

@article{arora2023spatial,
  title={Spatial transcriptomics reveals distinct and conserved tumor core and edge architectures that predict survival and targeted therapy response},
  author={Arora, Rohit and Cao, Christian and Kumar, Mehul and Sinha, Sarthak and Chanda, Ayan and McNeil, Reid and Samuel, Divya and Arora, Rahul K and Matthews, T Wayne and Chandarana, Shamir and others},
  journal={Nature communications},
  volume={14},
  number={1},
  pages={5029},
  year={2023},
  publisher={Nature Publishing Group UK London}
}

@article{bassiouni2023spatial,
  title={Spatial transcriptomic analysis of a diverse patient cohort reveals a conserved architecture in triple-negative breast cancer},
  author={Bassiouni, Rania and Idowu, Michael O and Gibbs, Lee D and Robila, Valentina and Grizzard, Pamela J and Webb, Michelle G and Song, Jiarong and Noriega, Ashley and Craig, David W and Carpten, John D},
  journal={Cancer research},
  volume={83},
  number={1},
  pages={34--48},
  year={2023},
  publisher={American Association for Cancer Research}
}

@article{berrocal2010spatio,
  title={A spatio-temporal downscaler for output from numerical models},
  author={Berrocal, Veronica J and Gelfand, Alan E and Holland, David M},
  journal={Journal of agricultural, biological, and environmental statistics},
  volume={15},
  number={2},
  pages={176--197},
  year={2010},
  publisher={Springer}
}

@article{cao_regression_2015,
	title = {Regression {Analysis} of {Sparse} {Asynchronous} {Longitudinal} {Data}},
	volume = {77},
	copyright = {https://academic.oup.com/journals/pages/open\_access/funder\_policies/chorus/standard\_publication\_model},
	issn = {1369-7412, 1467-9868},
	language = {en},
	number = {4},
	urldate = {2025-01-15},
	journal = {Journal of the Royal Statistical Society Series B: Statistical Methodology},
	author = {Cao, Hongyuan and Zeng, Donglin and Fine, Jason P.},
	month = sep,
	year = {2015},
	pages = {755--776},
}

@article{carroll_analysis_1970,
	title = {Analysis of {Individual} {Differences} in {Multidimensional} {Scaling} {Via} an {N}-way {Generalization} of “{Eckart}-{Young}” {Decomposition}},
	volume = {35},
	copyright = {https://www.cambridge.org/core/terms},
	issn = {0033-3123, 1860-0980},
	language = {en},
	number = {3},
	urldate = {2025-01-20},
	journal = {Psychometrika},
	author = {Carroll, J. Douglas and Chang, Jih-Jie},
	month = sep,
	year = {1970},
	pages = {283--319},
}

@article{chen2015spatially,
  title={Spatially resolved, highly multiplexed RNA profiling in single cells},
  author={Chen, Kok Hao and Boettiger, Alistair N and Moffitt, Jeffrey R and Wang, Siyuan and Zhuang, Xiaowei},
  journal={Science},
  volume={348},
  number={6233},
  pages={aaa6090},
  year={2015},
  publisher={American Association for the Advancement of Science}
}

@article{chen2022spatially,
  title={Spatially resolved transcriptomics reveals genes associated with the vulnerability of middle temporal gyrus in Alzheimer’s disease},
  author={Chen, Shuo and Chang, Yuzhou and Li, Liangping and Acosta, Diana and Li, Yang and Guo, Qi and Wang, Cankun and Turkes, Emir and Morrison, Cody and Julian, Dominic and others},
  journal={Acta neuropathologica communications},
  volume={10},
  number={1},
  pages={188},
  year={2022},
  publisher={Springer}
}

@article{chen_spatial_2020,
	title = {Spatial {Transcriptomics} and {In} {Situ} {Sequencing} to {Study} {Alzheimer}’s {Disease}},
	volume = {182},
	issn = {00928674},
	doi = {10.1016/j.cell.2020.06.038},
	language = {en},
	number = {4},
	urldate = {2025-01-30},
	journal = {Cell},
	author = {Chen, Wei-Ting and Lu, Ashley and Craessaerts, Katleen and Pavie, Benjamin and Sala Frigerio, Carlo and Corthout, Nikky and Qian, Xiaoyan and Laláková, Jana and Kühnemund, Malte and Voytyuk, Iryna and Wolfs, Leen and Mancuso, Renzo and Salta, Evgenia and Balusu, Sriram and Snellinx, An and Munck, Sebastian and Jurek, Aleksandra and Fernandez Navarro, Jose and Saido, Takaomi C. and Huitinga, Inge and Lundeberg, Joakim and Fiers, Mark and De Strooper, Bart},
	month = aug,
	year = {2020},
	pages = {976--991.e19},
}

@article{de2016cellular,
  title={The cellular phase of Alzheimer’s disease},
  author={De Strooper, Bart and Karran, Eric},
  journal={Cell},
  volume={164},
  number={4},
  pages={603--615},
  year={2016},
  publisher={Elsevier}
}

@article{efron2004least,
  title={Least Angle Regression},
  author={Efron, Bradley and Hastie, Trevor and Johnstone, Iain and Tibshirani, Robert},
  journal={Annals of Statistics},
  pages={407--451},
  year={2004},
  publisher={JSTOR}
}

@book{fan2018local,
  title={Local polynomial modelling and its applications: monographs on statistics and applied probability 66},
  author={Fan, Jianqing},
  year={2018},
  publisher={Routledge}
}

@article{fotheringham2017multiscale,
  title={Multiscale geographically weighted regression (MGWR)},
  author={Fotheringham, A Stewart and Yang, Wenbai and Kang, Wei},
  journal={Annals of the American Association of Geographers},
  volume={107},
  number={6},
  pages={1247--1265},
  year={2017},
  publisher={Taylor \& Francis}
}

@article{gotway2002combining,
  title={Combining incompatible spatial data},
  author={Gotway, Carol A and Young, Linda J},
  journal={Journal of the American Statistical Association},
  volume={97},
  number={458},
  pages={632--648},
  year={2002},
  publisher={Taylor \& Francis}
}

@article{hampel2021amyloid,
  title={The amyloid-$\beta$ pathway in Alzheimer’s disease},
  author={Hampel, Harald and Hardy, John and Blennow, Kaj and Chen, Christopher and Perry, George and Kim, Seung Hyun and Villemagne, Victor L and Aisen, Paul and Vendruscolo, Michele and Iwatsubo, Takeshi and others},
  journal={Molecular psychiatry},
  volume={26},
  number={10},
  pages={5481--5503},
  year={2021},
  publisher={Nature Publishing Group UK London}
}

@inproceedings{harshman_foundations_1970,
	title = {Foundations of the {PARAFAC} procedure: {Models} and conditions for an "explanatory" multi-model factor analysis},
	author = {Harshman, Richard A.},
	year = {1970},
}

@article{jack2018nia,
  title={NIA-AA research framework: toward a biological definition of Alzheimer's disease},
  author={Jack Jr, Clifford R and Bennett, David A and Blennow, Kaj and Carrillo, Maria C and Dunn, Billy and Haeberlein, Samantha Budd and Holtzman, David M and Jagust, William and Jessen, Frank and Karlawish, Jason and others},
  journal={Alzheimer's \& dementia},
  volume={14},
  number={4},
  pages={535--562},
  year={2018},
  publisher={Wiley Online Library}
}

@article{jack2024revised,
  title={Revised criteria for diagnosis and staging of Alzheimer's disease: Alzheimer's Association Workgroup},
  author={Jack Jr, Clifford R and Andrews, J Scott and Beach, Thomas G and Buracchio, Teresa and Dunn, Billy and Graf, Ana and Hansson, Oskar and Ho, Carole and Jagust, William and McDade, Eric and others},
  journal={Alzheimer's \& Dementia},
  volume={20},
  number={8},
  pages={5143--5169},
  year={2024},
  publisher={Wiley Online Library}
}

@article{jin2024advances,
  title={Advances in spatial transcriptomics and its applications in cancer research},
  author={Jin, Yang and Zuo, Yuanli and Li, Gang and Liu, Wenrong and Pan, Yitong and Fan, Ting and Fu, Xin and Yao, Xiaojun and Peng, Yong},
  journal={Molecular Cancer},
  volume={23},
  number={1},
  pages={129},
  year={2024},
  publisher={Springer}
}

@article{kiers_towards_2000,
	title = {Towards a standardized notation and terminology in multiway analysis},
	volume = {14},
	copyright = {http://doi.wiley.com/10.1002/tdm\_license\_1.1},
	issn = {0886-9383, 1099-128X},
	doi = {10.1002/1099-128X(200005/06)14:3<105::AID-CEM582>3.0.CO;2-I},
	language = {en},
	number = {3},
	urldate = {2025-01-20},
	journal = {Journal of Chemometrics},
	author = {Kiers, Henk A. L.},
	month = may,
	year = {2000},
	pages = {105--122},
}

@article{kolda_tensor_2009,
	title = {Tensor {Decompositions} and {Applications}},
	volume = {51},
	issn = {0036-1445, 1095-7200},
	doi = {10.1137/07070111X},
	language = {en},
	number = {3},
	urldate = {2025-09-16},
	journal = {SIAM Review},
	author = {Kolda, Tamara G. and Bader, Brett W.},
	month = aug,
	year = {2009},
	pages = {455--500},
}

@incollection{kruskal1989rank,
  title={Rank, decomposition, and uniqueness for 3-way and N-way arrays},
  author={Kruskal, Joseph B},
  booktitle={Multiway data analysis},
  pages={7--18},
  year={1989}
}

@article{li_asynchronous_2023,
	title = {Asynchronous {Functional} {Linear} {Regression} {Models} for {Longitudinal} {Data} in {Reproducing} {Kernel} {Hilbert} {Space}},
	volume = {79},
	copyright = {https://academic.oup.com/journals/pages/open\_access/funder\_policies/chorus/standard\_publication\_model},
	issn = {0006-341X, 1541-0420},
	language = {en},
	number = {3},
	urldate = {2025-02-27},
	journal = {Biometrics},
	author = {Li, Ting and Zhu, Huichen and Li, Tengfei and Zhu, Hongtu},
	month = sep,
	year = {2023},
	pages = {1880--1895},
}

@article{li_regression_2022,
	title = {Regression {Analysis} of {Asynchronous} {Longitudinal} {Functional} and {Scalar} {Data}},
	volume = {117},
	issn = {0162-1459, 1537-274X},
	language = {en},
	number = {539},
	urldate = {2025-02-27},
	journal = {Journal of the American Statistical Association},
	author = {Li, Ting and Li, Tengfei and Zhu, Zhongyi and Zhu, Hongtu},
	month = jul,
	year = {2022},
	pages = {1228--1242},
}

@article{mallach2024microglia,
  title={Microglia-astrocyte crosstalk in the amyloid plaque niche of an Alzheimer’s disease mouse model, as revealed by spatial transcriptomics},
  author={Mallach, Anna and Zielonka, Magdalena and van Lieshout, Veerle and An, Yanru and Khoo, Jia Hui and Vanheusden, Marisa and Chen, Wei-Ting and Moechars, Daan and Arancibia-Carcamo, I Lorena and Fiers, Mark and others},
  journal={Cell Reports},
  volume={43},
  number={6},
  year={2024},
  publisher={Elsevier}
}

@article{maynard2021transcriptome,
  title={Transcriptome-scale spatial gene expression in the human dorsolateral prefrontal cortex},
  author={Maynard, Kristen R and Collado-Torres, Leonardo and Weber, Lukas M and Uytingco, Cedric and Barry, Brianna K and Williams, Stephen R and Catallini, Joseph L and Tran, Matthew N and Besich, Zachary and Tippani, Madhavi and others},
  journal={Nature neuroscience},
  volume={24},
  number={3},
  pages={425--436},
  year={2021},
  publisher={Nature Publishing Group US New York}
}

@article{miyoshi2024spatial,
  title={Spatial and single-nucleus transcriptomic analysis of genetic and sporadic forms of Alzheimer’s disease},
  author={Miyoshi, Emily and Morabito, Samuel and Henningfield, Caden M and Das, Sudeshna and Rahimzadeh, Negin and Shabestari, Sepideh Kiani and Michael, Neethu and Emerson, Nora and Reese, Fairlie and Shi, Zechuan and others},
  journal={Nature Genetics},
  volume={56},
  number={12},
  pages={2704--2717},
  year={2024},
  publisher={Nature Publishing Group US New York}
}

@article{nadaraya1964estimating,
  title={On estimating regression},
  author={Nadaraya, Elizbar A},
  journal={Theory of Probability \& Its Applications},
  volume={9},
  number={1},
  pages={141--142},
  year={1964},
  publisher={SIAM}
}

@article{ni2022spotclean,
  title={SpotClean adjusts for spot swapping in spatial transcriptomics data},
  author={Ni, Zijian and Prasad, Aman and Chen, Shuyang and Halberg, Richard B and Arkin, Lisa M and Drolet, Beth A and Newton, Michael A and Kendziorski, Christina},
  journal={Nature Communications},
  volume={13},
  number={1},
  pages={2971},
  year={2022},
  publisher={Nature Publishing Group UK London}
}

@article{oshan2019mgwr,
  title={mgwr: A Python implementation of multiscale geographically weighted regression for investigating process spatial heterogeneity and scale},
  author={Oshan, Taylor M and Li, Ziqi and Kang, Wei and Wolf, Levi J and Fotheringham, A Stewart},
  journal={ISPRS International Journal of Geo-Information},
  volume={8},
  number={6},
  pages={269},
  year={2019},
  publisher={MDPI}
}

@article{shah2016situ,
  title={In situ transcription profiling of single cells reveals spatial organization of cells in the mouse hippocampus},
  author={Shah, Sheel and Lubeck, Eric and Zhou, Wen and Cai, Long},
  journal={Neuron},
  volume={92},
  number={2},
  pages={342--357},
  year={2016},
  publisher={Elsevier}
}

@article{staahl2016visualization,
  title={Visualization and analysis of gene expression in tissue sections by spatial transcriptomics},
  author={St{\aa}hl, Patrik L and Salm{\'e}n, Fredrik and Vickovic, Sanja and Lundmark, Anna and Navarro, Jos{\'e} Fern{\'a}ndez and Magnusson, Jens and Giacomello, Stefania and Asp, Michaela and Westholm, Jakub O and Huss, Mikael and others},
  journal={Science},
  volume={353},
  number={6294},
  pages={78--82},
  year={2016},
  publisher={American Association for the Advancement of Science}
}

@article{szpiro2013measurement,
  title={Measurement error in two-stage analyses, with application to air pollution epidemiology},
  author={Szpiro, Adam A and Paciorek, Christopher J},
  journal={Environmetrics},
  volume={24},
  number={8},
  pages={501--517},
  year={2013},
  publisher={Wiley Online Library}
}

@article{tucker_mathematical_1966,
	title = {Some {Mathematical} {Notes} on {Three}-{Mode} {Factor} {Analysis}},
	volume = {31},
	copyright = {https://www.cambridge.org/core/terms},
	issn = {0033-3123, 1860-0980},
	language = {en},
	number = {3},
	urldate = {2025-01-20},
	journal = {Psychometrika},
	author = {Tucker, Ledyard R},
	month = sep,
	year = {1966},
	pages = {279--311},
}

@article{vickovic2019high,
  title={High-definition spatial transcriptomics for in situ tissue profiling},
  author={Vickovic, Sanja and Eraslan, G{\"o}kcen and Salm{\'e}n, Fredrik and Klughammer, Johanna and Stenbeck, Linnea and Schapiro, Denis and {\"A}ij{\"o}, Tarmo and Bonneau, Richard and Bergenstr{\aa}hle, Ludvig and Navarro, Jos{\'e} Fernand{\'e}z and others},
  journal={Nature methods},
  volume={16},
  number={10},
  pages={987--990},
  year={2019},
  publisher={Nature Publishing Group US New York}
}

@article{wang2018three,
  title={Three-dimensional intact-tissue sequencing of single-cell transcriptional states},
  author={Wang, Xiao and Allen, William E and Wright, Matthew A and Sylwestrak, Emily L and Samusik, Nikolay and Vesuna, Sam and Evans, Kathryn and Liu, Cindy and Ramakrishnan, Charu and Liu, Jia and others},
  journal={Science},
  volume={361},
  number={6400},
  pages={eaat5691},
  year={2018},
  publisher={American Association for the Advancement of Science}
}

@article{xun2023reconstruction,
  title={Reconstruction of the tumor spatial microenvironment along the malignant-boundary-nonmalignant axis},
  author={Xun, Zhenzhen and Ding, Xinyu and Zhang, Yao and Zhang, Benyan and Lai, Shujing and Zou, Duowu and Zheng, Junke and Chen, Guoqiang and Su, Bing and Han, Leng and others},
  journal={Nature Communications},
  volume={14},
  number={1},
  pages={933},
  year={2023},
  publisher={Nature Publishing Group UK London}
}

@article{zeng_integrative_2023,
	title = {Integrative in situ mapping of single-cell transcriptional states and tissue histopathology in a mouse model of {Alzheimer}’s disease},
	issn = {1097-6256, 1546-1726},
	language = {en},
	urldate = {2025-01-15},
	journal = {Nature Neuroscience},
	author = {Zeng, Hu and Huang, Jiahao and Zhou, Haowen and Meilandt, William J. and Dejanovic, Borislav and Zhou, Yiming and Bohlen, Christopher J. and Lee, Seung-Hye and Ren, Jingyi and Liu, Albert and Tang, Zefang and Sheng, Hao and Liu, Jia and Sheng, Morgan and Wang, Xiao},
	month = feb,
	year = {2023},
}

@article{zhu2023srtsim,
  title={SRTsim: spatial pattern preserving simulations for spatially resolved transcriptomics},
  author={Zhu, Jiaqiang and Shang, Lulu and Zhou, Xiang},
  journal={Genome biology},
  volume={24},
  number={1},
  pages={39},
  year={2023},
  publisher={Springer}
}

@Manual{R-geoR,
  title        = {geoR: Analysis of Geostatistical Data},
  author       = {Ribeiro Jr, Paulo J. and Diggle, Peter J.},
  year         = {2025},
  note         = {R package version 1.9-6},
  doi          = {10.32614/CRAN.package.geoR},
}

@article{hong2016complement,
  title={Complement and microglia mediate early synapse loss in Alzheimer mouse models},
  author={Hong, Soyon and Beja-Glasser, Victoria F and Nfonoyim, Bianca M and Frouin, Arnaud and Li, Shaomin and Ramakrishnan, Saranya and Merry, Katherine M and Shi, Qiaoqiao and Rosenthal, Arnon and Barres, Ben A and others},
  journal={Science},
  volume={352},
  number={6286},
  pages={712--716},
  year={2016},
  publisher={American Association for the Advancement of Science}
}

@article{nasrabady2018white,
  title={White matter changes in Alzheimer’s disease: a focus on myelin and oligodendrocytes},
  author={Nasrabady, Sara E and Rizvi, Batool and Goldman, James E and Brickman, Adam M},
  journal={Acta neuropathologica communications},
  volume={6},
  number={1},
  pages={22},
  year={2018},
  publisher={Springer}
}

@article{mckenzie2017multiscale,
  title={Multiscale network modeling of oligodendrocytes reveals molecular components of myelin dysregulation in Alzheimer’s disease},
  author={McKenzie, Andrew T and Moyon, Sarah and Wang, Minghui and Katsyv, Igor and Song, Won-Min and Zhou, Xianxiao and Dammer, Eric B and Duong, Duc M and Aaker, Joshua and Zhao, Yongzhong and others},
  journal={Molecular neurodegeneration},
  volume={12},
  number={1},
  pages={82},
  year={2017},
  publisher={Springer}
}

@inproceedings{satopaa2011finding,
  title={Finding a" kneedle" in a haystack: Detecting knee points in system behavior},
  author={Satopaa, Ville and Albrecht, Jeannie and Irwin, David and Raghavan, Barath},
  booktitle={2011 31st international conference on distributed computing systems workshops},
  pages={166--171},
  year={2011},
  organization={IEEE}
}

@article{long2019alzheimer,
  title={Alzheimer disease: an update on pathobiology and treatment strategies},
  author={Long, Justin M and Holtzman, David M},
  journal={Cell},
  volume={179},
  number={2},
  pages={312--339},
  year={2019},
  publisher={Elsevier}
}

@article{Svensson2018SpatialDE,
  author  = {Svensson, Valentine and Teichmann, Sarah A. and Stegle, Oliver},
  title   = {{SpatialDE}: Identification of Spatially Variable Genes},
  journal = {Nature Methods},
  year    = {2018},
  volume  = {15},
  number  = {5},
  pages   = {343--346},
  doi     = {10.1038/nmeth.4636}
}

@article{Hu2021SpaGCN,
  author  = {Hu, Jian and Li, Xiangjie and Coleman, Kyle and Schroeder, Amelia and Ma, Nan and Irwin, David J. and Lee, Edward B. and Shinohara, Russell T. and Li, Mingyao},
  title   = {{SpaGCN}: Integrating Gene Expression, Spatial Location and Histology to Identify Spatial Domains and Spatially Variable Genes by Graph Convolutional Network},
  journal = {Nature Methods},
  year    = {2021},
  volume  = {18},
  pages   = {1342--1351},
  doi     = {10.1038/s41592-021-01255-8}
}

@article{Tanevski2022MISTy,
  author  = {Tanevski, Jovan and Flores, Ricardo O. R. and Gabor, Attila and Schapiro, Denis and Saez-Rodriguez, Julio},
  title   = {Explainable Multiview Framework for Dissecting Spatial Relationships from Highly Multiplexed Data},
  journal = {Genome Biology},
  year    = {2022},
  volume  = {23},
  pages   = {97},
  doi     = {10.1186/s13059-022-02663-5}
}

@article{Dong2022STAGATE,
  author  = {Dong, Kangning and Zhang, Shihua},
  title   = {Deciphering Spatial Domains from Spatially Resolved Transcriptomics with an Adaptive Graph Attention Auto-Encoder},
  journal = {Nature Communications},
  year    = {2022},
  volume  = {13},
  pages   = {1739},
  doi     = {10.1038/s41467-022-29439-6}
}

@article{Long2023GraphST,
  author  = {Long, Yuanhua and Ang, Kim Swee and Sethi, Ria and Ouyang, Jiarui F. and Li, Meng and Poletti, Giovanni and Ye, Chenxi J.},
  title   = {Spatially Informed Clustering, Integration, and Deconvolution of Spatial Transcriptomics with {GraphST}},
  journal = {Nature Communications},
  year    = {2023},
  volume  = {14},
  pages   = {1155},
  doi     = {10.1038/s41467-023-36796-3}
}

@article{Zhang2022STACI,
  author  = {Zhang, Xinyi and Wang, Xiao and Shivashankar, G. V. and Uhler, Caroline},
  title   = {Graph-Based Autoencoder Integrates Spatial Transcriptomics with Chromatin Images and Identifies Joint Biomarkers for {Alzheimer}'s Disease},
  journal = {Nature Communications},
  year    = {2022},
  volume  = {13},
  pages   = {7480},
  doi     = {10.1038/s41467-022-35233-1}
}

@inproceedings{salvador2004determining,
  title={Determining the number of clusters/segments in hierarchical clustering/segmentation algorithms},
  author={Salvador, Stan and Chan, Philip},
  booktitle={16th IEEE international conference on tools with artificial intelligence},
  pages={576--584},
  year={2004},
  organization={IEEE}
}

@article{fan1996study,
  title={A study of variable bandwidth selection for local polynomial regression},
  author={Fan, Jianqing and Gijbels, Ir{\`e}ne and Hu, Tien-Chung and Huang, Li-Shan},
  journal={Statistica Sinica},
  pages={113--127},
  year={1996},
  publisher={JSTOR}
}

@article{brunsdon2002geographically,
  title={Geographically weighted summary statistics—a framework for localised exploratory data analysis},
  author={Brunsdon, Chris and Fotheringham, AS and Charlton, Martin},
  journal={Computers, Environment and Urban Systems},
  volume={26},
  number={6},
  pages={501--524},
  year={2002},
  publisher={Elsevier}
}

@article{song2024scdesign3,
  title={scDesign3 generates realistic in silico data for multimodal single-cell and spatial omics},
  author={Song, Dongyuan and Wang, Qingyang and Yan, Guanao and Liu, Tianyang and Sun, Tianyi and Li, Jingyi Jessica},
  journal={Nature Biotechnology},
  volume={42},
  number={2},
  pages={247--252},
  year={2024},
  publisher={Nature Publishing Group US New York}
}

\end{document}